\documentclass[apj, twocolumn]{emulateapj}
\usepackage{epstopdf}
\newcommand{\bootes}{Bo\"otes\ } 
\newcommand{\oii}{[\ion{O}{2}]}

\begin{document}

\title{The PRIsm MUlti-object Survey (PRIMUS). II. Data Reduction and Redshift Fitting}
\author{
Richard J. Cool \altaffilmark{1}
John Moustakas \altaffilmark{2}
Michael R. Blanton \altaffilmark{3}
Scott M. Burles \altaffilmark{4}
Alison L. Coil \altaffilmark{5}
Daniel J. Eisenstein \altaffilmark{6}
Kenneth C. Wong \altaffilmark{7}
Guangtun Zhu \altaffilmark{8}
James Aird \altaffilmark{5}
Rebecca A. Bernstein \altaffilmark{9}
Adam S. Bolton \altaffilmark{10}
David W. Hogg \altaffilmark{3}
Alexander J. Mendez \altaffilmark{5}}
\altaffiltext{1}{MMT Observatory, Tucson AZ 85721}
\altaffiltext{2}{Department of Physics, Siena College, 515 Loudon Rd.,
Loudonville, NY 12211}
\altaffiltext{3}{Center for Cosmology and Particle Physics, Department of Physics, New York University, 4 Washington
Place, New York, NY 10003}
\altaffiltext{4}{D.E. Shaw \& Co. L.P, 20400 Stevens Creek Blvd., Suite 850, Cupertino, CA, 95014}
\altaffiltext{5}{Department of Physics, Center for Astrophysics and Space Sciences, University of California, 9500 Gilman Dr., La Jolla, San Diego, CA 92093}
\altaffiltext{6}{Harvard-Smithsonian center for Astrophysics, 60 Garden St, MS \#20, Cambridge, MA 02138}
\altaffiltext{7}{Steward Observatory, The University of Arizona, 933 N. Cherry Ave., Tucson, AZ 85721}
\altaffiltext{8}{Center for Astrophysical Sciences, Department of Physics
and Astronomy, Johns Hopkins University, 3400 North Charles
Street, Baltimore, MD 21218, USA.}
\altaffiltext{9}{Department of Astronomy and Astrophysics, UCA/Lick Observatory, 
University of California, 1156 High Street, Santa Cruz, CA 95064}
\altaffiltext{9}{Department of Physics and Astronomy, University of Utah, Salt Lake City, UT 84112}

\begin{abstract}

The PRIsm MUti-object Survey (PRIMUS) is a spectroscopic galaxy
redshift survey to $z\sim1$ completed with a low-dispersion prism and
slitmasks allowing for simultaneous observations of $\sim2,500$
objects over 0.18 deg$^2$. The final PRIMUS catalog includes $\sim$130,000
robust redshifts over 9.1deg$^2$. In this paper, we summarize the
PRIMUS observational strategy and present the data reduction details used to
measure redshifts, redshift precision, and survey completeness.
The survey motivation, observational techniques, fields, target
selection, slitmask design, and observations are presented in
\citet{Coil2010_PaperI}.  Comparisons to existing higher-resolution
spectroscopic measurements show a typical precision of
$\sigma_z/(1+z)=0.005$. PRIMUS, both in area and number of redshifts, is the largest
faint galaxy redshift survey completed to date and is allowing for precise 
measurements of the relationship between AGNs and their hosts, the
effects of environment on galaxy evolution, and the build up of galactic
systems over the latter half of cosmic history.

\end{abstract}

\section{Introduction}

The PRIsm MUlti-object Survey (PRIMUS) is a new spectroscopic survey of faint galaxies
focused on measuring the evolution of galaxy properties and large-scale structure since 
$z\sim1$.  PRIMUS uses a hybrid technique between very low-resolution redshifts
measured from galaxy photometry and high-resolution spectroscopy.  Rather than
using a more traditional grating or grism, we utilize a specially designed low-dispersion
prism to observe the full optical spectra of up to $\sim3,000$ objects simultaneously.  Combined
with the 0.2 deg$^2$ field of view provided by the IMACS spectrograph on the Baade I 6.5m telescope
at Las Campanas, PRIMUS spectroscopically measures galaxy redshifts in much less time than
would be required with more traditional high-resolution spectroscopic techniques ($\sim10,000$ galaxies per night) 
and reaches depths of $i\sim23$ in one hour.
PRIMUS redshifts have typical precision of $\sigma_z/(1+z)=0.005$.
PRIMUS spectroscopy focuses on fields with existing optical imaging for target selection and emphasizes 
areas with existing high-quality multiwavelength imaging from {\it GALEX}, {\it Spitzer}, {\it Chandra}, and
{\it XMM-Newton}.  The combination of PRIMUS redshifts with this wealth of multiwavelength data allows for detailed
measurements of galaxy colors, luminosities, star formation rates, and stellar masses. 

Traditional spectroscopic redshift surveys require considerable resources to be completed, especially 
as one probes galaxies beyond the local universe.  For example, the DEEP2 survey \citep{Davis:2003} 
required more than 80 nights on Keck and zCOSMOS \citep{Lilly:2007} and VVDS \citep{LeFevre:2005, Garilli:2008} 
utilized over one hundred VLT nights each.  With this large amount of observing time, these surveys 
probe $\sim20,000$ galaxies over a few square degrees. PRIMUS represents a new paradigm in galaxy redshift 
surveys which enables the efficient survey of large astronomical volumes with much less observation time 
than required with traditional higher-resolution spectroscopy.   The methods pioneered by PRIMUS have already 
been adopted and refined for other prism spectroscopic surveys of the high-redshift universe \citep{Kelson:2012, Just2012, Grayinprep}.

Science with the PRIMUS dataset is ongoing.  In \citet{Aird2012}, we measure 
X-ray observations in PRIMUS fields to find evidence the presence and accretion 
of AGNs does not depend on the stellar masses of their host galaxies. 
\citet{Wong2011} investigated tidally triggered star formation in isolated 
close galaxy pairs and measured an enhancement in the specific star formation 
rate due to tidal interactions in these galaxies.  We investigated optically
``red and dead'' galaxies which show signatures of star formation from their
infrared photometry in \citet{Zhu2011} and show that a significant fraction of
red-sequence galaxies have ongoing obscured star formation. In \citet{Moustakas2012}, we present the stellar mass function of PRIMUS galaxies and conclude that
star forming galaxies are quenched more strongly with decreasing stellar mass
and the majority of the stellar mass buildup within quiescent galaxies 
occurs around $\sim 10^{10.8} M_\odot$.

This paper describes the observations, data reduction, redshift fitting, redshift precision, and lessons learned
from our experience with low-resolution prism spectroscopy.  A companion paper \citep{Coil2010_PaperI} (hereafter Paper I)
describes the 
survey design and characteristics and provides an overview of the data taken as part of the PRIMUS survey.  The paper
is outlined as follow.  Section \ref{sec:obsstrat} outlines the details of PRIMUS observations including details of 
our nod \& shuffle spectroscopy, calibration data, and prism characteristics.   In \S\ref{sec:2dprocess}, we provide
a detailed discussion of our image reduction and spectral extraction including corrections due to scattered light and
our wavelength calibration.   We summarize our measurements of the flux corrections for our PRIMUS spectra in 
\S\ref{sec:1.5dcalibration}.  Section \ref{sec:1dfitting} introduces the spectral libraries we utilize to measure redshifts
from PRIMUS spectroscopy and detail the fitting algorithm.  The redshift precision and outlier rates are presented in \S\ref{sec:quality}
and completeness estimates are presented in \S\ref{sec:completeness} before concluding in \S\ref{sec:lessons}.  The first release of PRIMUS 
redshifts has been completed and can be found at http://primus.ucsd.edu.

\section{Observational Strategy}
\label{sec:obsstrat}

\begin{figure}
\includegraphics[width=3.5in]{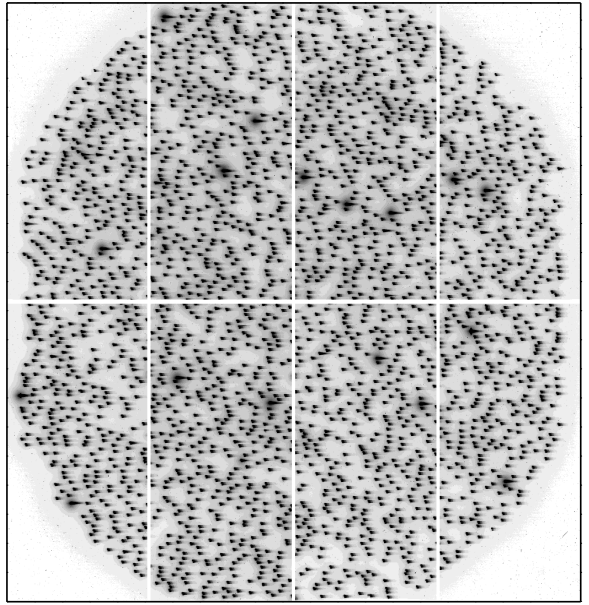}
\caption{\label{fig:fullimage} Mosaic of all eight IMACS CCDs for a single PRIMUS science exposure.  Each trace of light corresponds to
 two individual slits which have been shuffled on the detector, resulting in 4 traces per object. 
 The very bright traces in the image correspond to alignment stars
 utilized to ensure that the slitmask has been centered properly when pointing the telescope.}
\end{figure}

Traditional wide area multiobject spectrographic surveys utilize either slit masks or fiber-fed spectrographs to obtain 
spectra for tens or hundreds of galaxies simultaneously.  By replacing the traditional grism or grating dispersive 
element with a low dispersion prism, PRIMUS utilizes slit masks to obtain spectra for thousands of galaxies simultaneously.   
Figure \ref{fig:fullimage} shows a mosaic for all eight IMACS CCDs for one PRIMUS science exposure illustrating the 
large multiplexing power available with our low-dispersion prism technique. We design each PRIMUS slitmask to provide the maximum number 
of high priority
galaxies while also ensuring we dedicate enough slits on each mask for calibration purposes.  

PRIMUS utilized existing optical photometric catalogs for target selection and mask design.  Paper I details the galaxy and AGN
selection criteria, mask design, sparse sampling, and photometric catalog details used in our mask design.  Here, we focus on the
observational and analytical methods used to obtain spectra for PRIMUS targets.  

\begin{figure}
\includegraphics[width=3.5in]{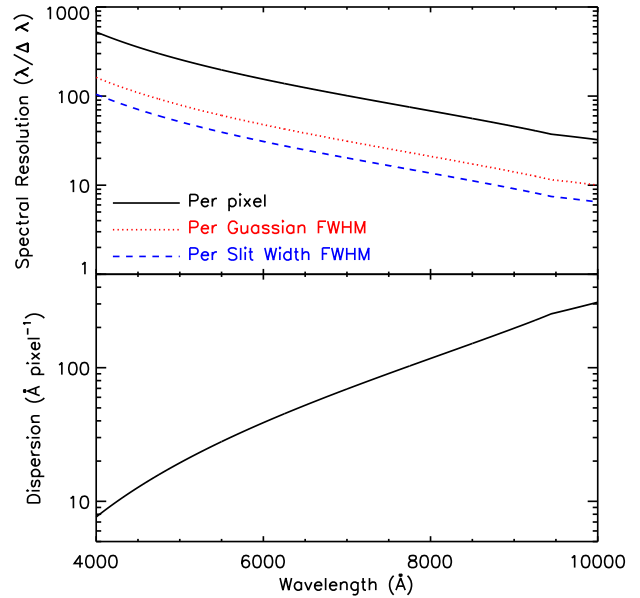}
\caption{\label{fig:primres} Resolution and
dispersion of the PRIMUS prism versus wavelength.  The resolution and
dispersion generated by the low dispersion prism is a strong function of the wavelength 
with increased resolution in the blue but low resolution at $9000\AA-1\mu m$.
The low resolution allows us to observe the full optical spectrum of a 
target galaxy on only $\sim$150 pixels on the detector
and up to  $\sim$3000 galaxies on a single mask. }
\end{figure}

All PRIMUS spectroscopy was observed in nod \& shuffle mode \citep{Glazebrook:2001} with IMACS on the Baade Magellan I telescope
at Las Campanas.  Figure \ref{fig:primres} shows
the PRIMUS prism resolution and dispersion as a function of wavelength.  The prism resolution is a strong function of wavelength
and has lowest resolution in the red. Robust sky subtraction is vital for PRIMUS redshift fitting as most galaxies are fit
based on the shape of the continuum (galaxies with strong emission lines include those lines in the redshift fitting, but many galaxies in our sample have weak or absent emission lines). As nod \& shuffle allows us to measure the sky flux in the same pixels as we measure object flux, 
we can perform robust sky subtraction.

\begin{figure}[!h]
\includegraphics[width=3in, angle=90]{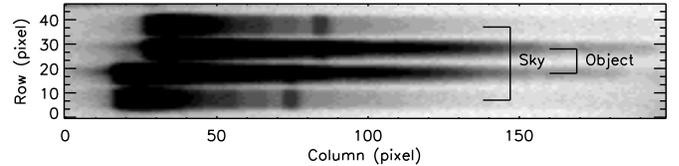}
\caption{\label{fig:exampletrace} Example
of a full bright-object trace from PRIMUS spectroscopy.  PRIMUS observations are 
completed in nod \& shuffle mode with the object nodded between two slits
separated in both right ascension and declination.  This pattern results in
four traces on the final image, two traces of pure sky light and two traces of
object combined with sky.  In this figure, red is to the left and blue to the 
right.  For reference, the bright sky line near pixel 73 (83) in the bottom (top)
trace corresponds to 5577 \AA.  The resolution and dispersion of the PRIMUS
prism are strong functions of wavelength; the forest of emission lines
in the red portion of the spectrum are blended at PRIMUS
resolution. }
\end{figure}

Rather than nodding the mask such that objects moved along one long slit, we adopted a strategy of cutting two slits (often referred to 
as slit A and slit B throughout this paper) 
for each object on the mask separated by $(\Delta\alpha, \Delta\delta)=(-1\farcs6, -3\farcs2)$ when we designed 0$\farcs$8 wide
slits and $(\Delta\alpha, \Delta\delta)=(-2\farcs0, -4\farcs0)$ when we designed 1$\farcs$0 slits.  By nodding in both directions
on the sky rather than in a single direction, we minimize the impact
of a bad column on the final extracted spectrum of an observed galaxy;
a galaxy moved in both directions will not lose all of the information from the wavelength associated with the bad column.  During
each telescope nod, the detector was also shuffled 8 (10) pixels for masks with 0$\farcs$8 (1$\farcs0$) wide slits.  An example of
nod \& shuffle traces for a single object is show in Figure \ref{fig:exampletrace}.  Due to nodding between two slits, the final
observation includes four traces for each object; two traces with only sky illumination and two with both sky and object contributions.
In November 2007, modifications to the IMACS instrument lead to a slight change in this layout; objects after this
date were observed in the outer two traces and the inner two traces contained only sky light.  Throughout this paper we will 
often discuss slits (either object slits or sky slits), referring to the spectrum observed through a single slit (in either object
or sky position). When we discuss the "full object trace" we refer to the collection of the 4 traces shown in Figure \ref{fig:exampletrace} 
including the 2 sky traces and 2 objects traces coming from slit A and slit B in nod \& shuffle mode.

Science frames were typically observed in four 16 minute nod \& shuffle exposures for a total exposure time of 64 minutes (32 minutes
in each nod \& shuffle object slit).  In COSMOS, we exposed deeper due to the wealth of high-quality data in the field;  COSMOS
masks included six 16 minute exposures for a total of 96 minutes per mask.  

\section{Image Analysis and Spectral Extraction}
\label{sec:2dprocess}

We utilize a multi-stage process to first remove contamination 
light from the the full two-dimensional science frames and then 
to extract the one-dimensional spectrum of each observed object.  
In this section, we provide the details relevant to our processing 
of IMACS images and the methods we utilize to obtain high-fidelity
low-dispersion spectra via nod \& shuffle sky subtraction. Finally, we discuss 
the modeling methods used to extract object spectra.

\subsection{Large Scale Scattered Light}

IMACS images suffer from large diffuse halos of light around bright sources.  The 
total light contained in this halo is correlated with the flux of nearby objects and
is extended enough to influence the light gathered in nearby object traces. The scattered light crosses chip gaps, 
indicating that the halos are not due to scattering within the CCDs themselves but 
are likely created in the camera optics.  The presence of excess light
will lead to errors in sky subtraction and can easily contaminate
extracted spectra if not carefully corrected.  

\begin{figure}
\includegraphics[width=2.75in, angle=90]{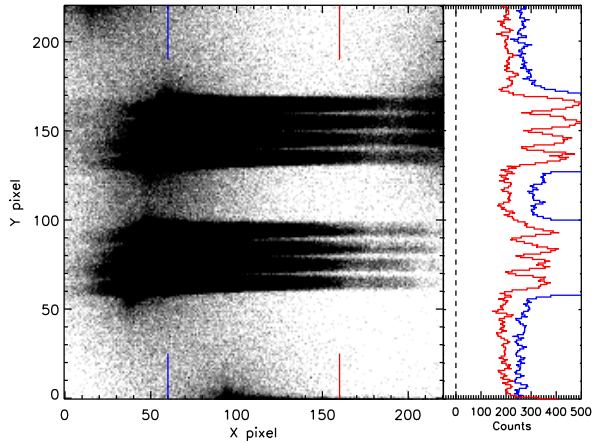}
\caption{\label{fig:haloraw} Left: Two dimensional subregion from one PRIMUS science exposure. Two object trace sets (each object
set is composed of four traces) are shown.  The presence of light between the two trace sets is the signature of the large scale 
scattered light halo we remove before performing object extraction from our data.  The red and blue vertical lines illustrate the
locations where two slices were made through the image.  The slices show that this scattered light contributes several 
hundred counts per pixel and is a clear contaminant to our spectra.  Furthermore, since the scattered light has two dimensional 
structure on our images, performing nod \& shuffle sky subtraction without removing the contaminating signal will result in poor sky subtraction.}
\end{figure}

Figure \ref{fig:haloraw} illustrates the presence of the large scale halo light near two object trace sets from a PRIMUS
science exposure.  On the right, we show two slices through the image; the red and blue slices are taken at the locations
of the vertical red and blues lines in the two-dimensional image.  The clear signal of light between the trace sets is the signature
of the large scale scattering halo we characterize and remove using the details in this subsection.

In order to remove the contribution from the large-scale scattering light component, we first stack
two-dimensional images of full traces with no targeted object (originally milled to characterize sky emission)
and use this stacked image to create
a model for the scattered light profile. We parameterize the scattered light term as a convolution of 
our image with kernel, $k(x,y)$

\begin{equation}
k(x,y)=\left(1+\frac{x^2+y^2}{r_0^2}\right)^{-p}
\label{eq:halo}
\end{equation}
where $x$ and $y$ correspond to the pixel coordinates on the detector and $r$ and $p$ are free
parameters we fit to characterize the shape of the scattered light halo.

We create a two-dimensional model for the scattered light by coadding all of the science exposures for a given 
pointing to 
increase the signal-to-noise
ratio in the scattered light and remove cosmic rays.
We mask the location
of full traces on the coadded image 
 and then fit for the kernel scale length, $r_0$, and power-law index, $p$, over the full 0.2 deg$^2$ 
IMACS focal plane using the Levenberg-Marquardt method with the IDL routine {\tt mpfit} \citep{mpfit}.

We find that a simple convolution of the best-fit kernel and
the original science image is not sufficient to provide a full
correction to the scattered light due to several factors. 
First, the amplitude of the scattered light with
respect to the incident light from the slits varies as a function of
position on the focal plane. Second, we find that, in detail, the
large-scale halo is not circularly symmetric as assumed by equation~(\ref{eq:halo}).
In order to obtain a better model of the scattered
light, we define two non-circularly symmetric kernels based on the
best fitting $k(x,y)$

\begin{eqnarray}
\theta = {\rm tan}^{-1}(y/x) \\
k_1(x,y) = k(x,y)\, {\rm cos}(\theta) \\
k_2(x,y) = k(x,y)\, {\rm cos}(\theta+\pi/2).
\end{eqnarray}

We solve for the spatial dependence of the halo term by convolving the masked image with
the three kernels to generate three halo terms
\begin{eqnarray}
M_{\rm halo} = I \otimes k \\
D_{1,\rm halo} = I \otimes k_1 \\
D_{2,\rm halo} = I \otimes k_2.
\end{eqnarray}

\begin{figure}[b]
\includegraphics[width=2.75in, angle=90]{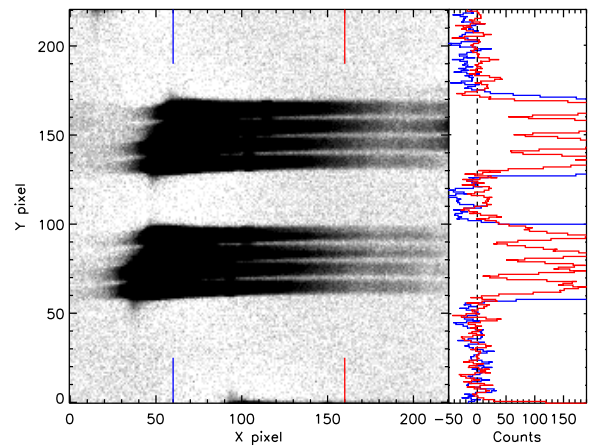}
\caption{\label{fig:halolargesubtract} Two dimensional subregion shown in Figure \ref{fig:haloraw} after the best fitting model 
for the large scale halo has been subtracted from the image. The slices through the image plotted on the right do not show signal between the traces.}
\end{figure}

We then solve for the spatial variation in the halo by creating the final halo image, 
$I_{\rm halo}$, and solving for $w_i$ to minimize the light between object traces
\begin{eqnarray}
I_{\rm halo}(x,y) = (w_0+ w_1x + w_2y + w_3x^2+w_4y^2) 	+ \\ 
     \nonumber M_{\rm halo}(w_5+w_6x+w_7y+w_8x^2+w_9y^2)  + \\
	 \nonumber w_{10}D_{1, {\rm halo}} + w_{11}D_{2, {\rm halo}}.
\end{eqnarray}
When applying the best fitting halo image, $I_{\rm halo}$, constructed in this manner, we normalize the counts in the
halo to the median counts in each individual exposure to adjust for exposure to exposure differences.  
Figure \ref{fig:halolargesubtract} follows Figure \ref{fig:haloraw} but illustrates the same
two dimensional image after the large scale halo model has been subtracted.  The slices shown on the right in the 
figure highlights the success of this technique; little light remains between the object traces.

\subsubsection{Small Scale Scattered Light}
\label{sec:smallscalehalo}
In addition to the large scale scattering halo, our IMACS spectroscopic images
have a second small-scale scattering component on  5--10 pixel scales.
This halo component 
is most apparent in the red portion 
of the spectra and is non-axisymmetric with respect to the light in the 
full traces.  On average, this small scale halo light
corresponds to approximately 1\% of the flux of the sky. 
While the cause of this excess light is 
unknown, the local nature of the light suggests it may be due to diffusion 
of red light in the CCDs.

\begin{figure}
\includegraphics[width=3in]{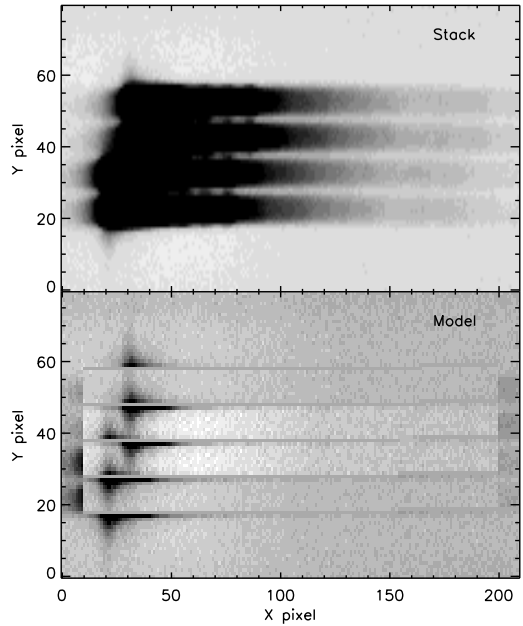}
\caption{\label{fig:empsubhalo2d} Top: Stacked image of all of the isolated traces on a single PRIMUS observation which illustrates the
presence of small-scale scattered light (for examples the ``spurs" of light extending above and below the traces).  Bottom: 
Empirically constructed model used to 
remove the small-scale scattered light constructed as described in \S\ref{sec:smallscalehalo}.}
\end{figure}

For each CCD, we first locate all full trace sets 
without another trace within 20 pixels. 
For each night of observations, we stack the
two dimensional image of these isolated traces after performing
a nod \& shuffle sky subtraction.  We then quantify the presence
of excess light outside of the sky traces in the composite image. We
assume the object trace experiences the same amount of scattering as
the sky trace (although mirrored across the trace) and construct
a final empirical model for the small scale scattered light in the
spectra.
In the top panel of 
Figure \ref{fig:empsubhalo2d}, we show an example of such a stacked image. 
The scattered light is apparent especially in the red (left side).
We take the
scattering light in the 7 (9) pixels (for $0.8''$ ($1''.0$) wide slits)
above the top trace and that in the 7 (9) pixels below the bottom trace, 
and reproduce the light for each trace based on the same nod \& shuffle 
mode as observations. In the bottom panel of Figure \ref{fig:empsubhalo2d},
we show the empirical model constructed from the stacked image in the top panel. 
Finally, we duplicate this model for each trace on the CCD to subtract 
the scattering light from the image. Overall, this correction is $\sim1\%$ of the sky and thus
an important contamination to remove from our data before extracting the one-dimensional spectra.

\subsection{Flatfield Correction}

At this point, we have performed no flat fielding of the PRIMUS two-dimensional
images.  When fitting
PRIMUS spectra to obtain redshifts, we are not concerned with the normalization
of the spectra (which is affected by vignetting and slit losses) but
removal of pixel-to-pixel changes from the detector is vital in order to not
introduce false spectral features in our extracted spectra.  When constructing a 
flat field correction for our data, we first remove any large-scale gradients
present in our observed twilight sky flats.  The resulting flat field measures
the pixel-to-pixel variations across the detector.  Since the twilight flat is observed
as a single image and our science frames are observed with nod \& shuffle, we cannot simply
divide the science data by this pixel-to-pixel flat field correction.

Using our normalized flat-field, we construct
a ``nod \& shuffle'' corrected flat field by finding the mean between the 
observed flat field image and one shifted by our shuffle length.  In general, 
PRIMUS spectra are sky-dominated, so the assumption that each nod position 
contributes evenly to the total contribution to the flat field is sufficient
to correct all but the most severe blemishes in the flat field.  Areas on the
detector that are affected by flat-field corrections of more than 25\% are
masked as possible bad pixels and are ignored throughout the redshift fitting
process to ensure that these large flat-field corrections, where our simple
assumption may not be ideal, do not affect our final spectra.  Less than 
1\% of the pixels are masked to this large flat fielding error.   This final 
flat frame is utilized in our forward modeling of the PRIMUS science frames; 
in contrast to more traditionally approaches, we do not directly apply the
flat field correction to our images.

\subsection{Spectral Modeling \& Extraction}
Having corrected for the scattered light in the two dimensional images,
we are ready to extract the one-dimensional object spectra from the images.  In PRIMUS, we utilize
a forward modeling approach to spectral extraction; we first perform an extraction without information
on the spatial profile of each object.  We then perform more robust extractions to construct a 
one dimensional profile and perform an extraction based on that profile to obtain the final 
extracted spectra.  This section details the steps involved with calibrating the two
dimensional images and extract science-quality spectra.

\subsubsection{First pass extraction}
We perform spectral extraction of the halo-corrected two dimensional PRIMUS
spectra in a three pass method.  The first pass of extraction allows us to 
derive the spatial profile used to extract each object, the second pass 
refines the sky subtraction and the final pass measures the one-dimensional
extracted spectrum for each object. 

In the first pass, we use a simple model for the 
sky and object light in order to build a more detailed spatial profile for
each object in the subsequent extraction passes. Extraction is performed in a 
forward-modeling sense on a column-by-column basis in the images.  For each
column, we identify pixels associated with each PRIMUS trace.  
We construct models, $S_i$, the sky light in slit
$i$ (either slit A or slit B), and $O_i$, the object light in either slit 
position.  In the first pass extraction, we construct simple models for each
\begin{equation}
S_i = \sum_j^{N_{\rm pix}} w_{s,j} \delta_j f_j
\end{equation}
\begin{equation}
\label{eqn:firstpass}
O_i = \sum_j^{N_{\rm pix}} w_{o,j} \delta_j f_j
\end{equation}
where $f_j$ is the flat field contribution to pixel $j$ in the trace, 
$\delta_j$ are delta-functions in each of the pixels across the trace, and $w_{s,j}$ and 
$w_{o,j}$ are the fluxes measured in sky and object pixels prior to the 
flat-field contribution.  Before
fitting, we construct a model for the full four trace contribution to the
column from the object; the sky contributes flux only in the sky slits and both 
the sky and object model contribute flux in the object traces.  Finally, this
model is convolved with a boxcar with the width of the mask slit and a 
Gaussian with a 1.5 pixel width to represent the instrumental profile.
By solving for $w_{s,j}$ and $w_{o,j}$, we construct a rough model
for the sky and object illumination in the slit.

\subsubsection{Second and third pass extractions}
The second pass of extraction uses the output from the first pass
in order to more robustly extract the object spectra.  We start this process
by generating a spatial profile of the object light in each slit.  We sum 
the object light, as measured in the first pass, along the spectral direction
of the PRIMUS trace to measure $p_j$, the total
light in the slit as a function of pixel $j$ across the slit. We do this
sum separately for each A and B nod \& shuffle position; in an ideal
observation, these spatial profiles would be identical, but due to guiding 
errors or errors in the mask cutting, the objects may not be centered
at the same pixel in both positions.

We fit $p_j$ with a Gaussian model (solving for the centroid
and Gaussian width) in order to model the spatial profile of the object. While
this measurement is robust for well-detected objects in a single exposure, 
the lowest signal-to-noise objects have derived profiles that are heavily 
influenced by noise in the images.  Rather than using noisy profiles for the
low signal-to-noise objects, we force these objects to have the median width
and centroid of the other objects on the same CCD.  Based on the inferred 
centroid, $\mu_i$, and profile width, $\sigma_i$, of object $i$, we construct
a new object model
\begin{equation}
O_{i,j} = w_{i,j} {\rm exp}\left(-\frac{(j-\mu)^2}{\sigma^2}\right) f_j.
\end{equation}

\begin{figure}
\includegraphics[width=3in]{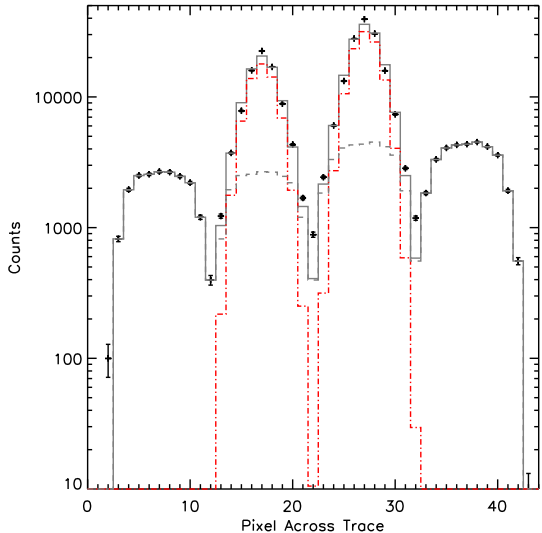}
\caption{\label{fig:secondpasscut} Results of the second pass of PRIMUS spectral extraction.  The black data points with error bars shows 
the actual PRIMUS data while the red (dot-dashed) and grey denote the best fitting object and sky contributions to the four traces. The final combined model
is shown by the grey line.}
\end{figure}

We perform the extraction column-by-column using this spatial profile
for the object traces rather than the $\delta$-function formulation
in equation \ref{eqn:firstpass}.  Figure \ref{fig:secondpasscut} shows 
a slice through one relatively bright PRIMUS object trace set.  The red lines show the
best fitting object profile while the blue shows the contribution from the sky.  The grey line
shows the sum of the sky and object models compared to the black data points with errors showing the actual PRIMUS
data.   We utilize the best fitting
sky parameters from this pass to create a two-dimensional sky image
which is subtracted from the halo-corrected two-dimensional science frame.
We perform a final extraction utilizing the spatial profile derived above
to extract each object trace with an optimal extraction 
algorithm \citep{Horne:1986} which maximizes the signal-to-noise 
ratio in the extracted spectra compared to a simple boxcar
extraction.

\begin{figure}
	\includegraphics[width=3in]{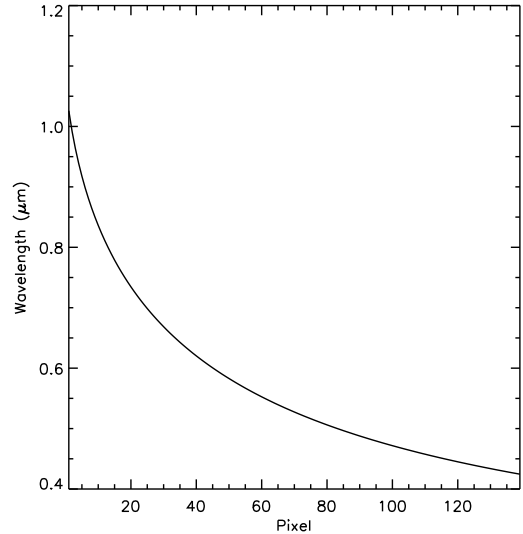}
	\caption{\label{fig:wavevector} Wavelength as a function of pixel position along the trace for PRIMUS observations.  This 
	characteristic curve has pixels roughly evenly spaced in $\lambda^{-3}$.  When solving for the final wavelength vector for
	 each object in the survey, we measure the zeropoint offset between this fiducial model and the observations based on the 
	locations of key sky emission features observed in each sky slit in our nod \& shuffle exposures.}
\end{figure}

\subsection{Wavelength calibration}
\begin{figure}
	\includegraphics[width=3in]{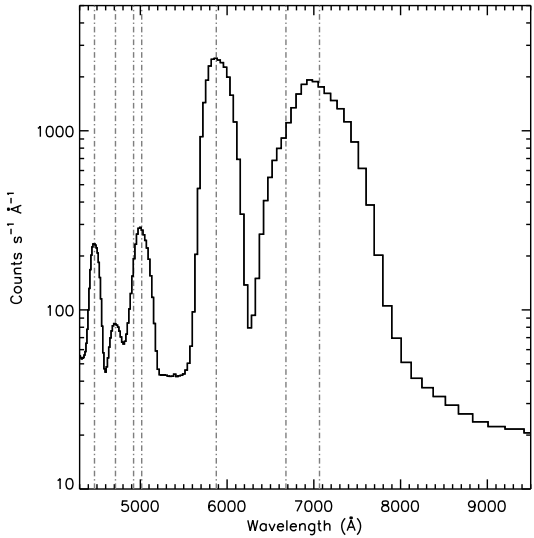}
	\caption{\label{fig:henear} Comparison of the wavelength solution determined by using an optical model for the IMACS optical path 
	combined with our refinement based
	on the location of prominent features in the sky spectrum with an extraction of a helium lamp observation through the same slit.  
	Using a technique based on sky line features allows us to determine wavelength corrections based on simultaneous observations with 
	our object spectra rather than extrapolating corrections from images which bracket our science frames.  The location of key lines 
	in the helium spectrum are highlighted in by dot-dashed lines.}
\end{figure}

We perform wavelength calibration using a three step process.  Initially,
we apply a model wavelength vector to each extracted object and sky 
spectrum based on the optical model for the IMACS spectrograph and
PRIMUS prism.  In this model, the pixels are spaced roughly evenly in  
$\lambda^{-3}$. Figure \ref{fig:wavevector} shows the wavelength as a function of  pixel
from this optical model.
In order to ensure that this model correctly constrains the final performance
of the prism, we observed images illuminated by helium arc lamps on the flat
field screen.  We perform a
simple extraction of the arc images and compare the spacing of the
well-separated helium lines with the expected positions based on the prism
dispersion model and find excellent agreement for all PRIMUS observing runs.
Figure \ref{fig:henear} shows one such comparison; throughout all runs, the general 
form of the relationship between pixel and wavelength remained fixed.  

While the optical model of the instrument and prism perform well in
predicting the dispersion in PRIMUS spectra, it cannot accurately predict
the zeropoint for the wavelength solution on the scales needed to achievement
robust redshifts.  While the PRIMUS spectra are low resolution, the brightest
sky lines are still readily detected and make ideal zeropoint normalizations
to our wavelength solutions.  Based on the strong night sky lines at 5577.34, 
5895.00, 6300.3 $\AA$, we
derive a zeropoint offset for each slit on a CCD.  Due to the low resolution nature of
our spectra, we find the most robust zeropoints are determined by doing a cross correlation
between a model sky spectrum and our observed sky in windows around each of the bright sky lines.  
Based on this cross correlation, we measure an offset for each object on a given PRIMUS
exposure. In order to
properly calibrate objects where a portion of the bright sky lines are 
affected by bad columns and to protect against contamination from cosmic rays
which may occur if we apply a slit-by-slit zeropoint correction, we fit a low 
order polynomial to the 
the final zeropoint shifts as a function of position on the IMACS focal plane.
We then apply the corresponding model value to each slit in our science 
observation.

The zeropoint derived from the sky lines is appropriate if the slit is filled evenly by the object
or for slits in which the object is centered.  Due
to alignment errors, mask cutting errors, astrometry errors in the photometric catalog, etc, 
this assumption is not always true for PRIMUS observations.  Ignoring the effect
of non-aligned slits can lead to systematic errors in our final wavelength calibration, and thus
we utilize the $\sim30$ F-stars observed
on each PRIMUS mask in order to derive systematic 
zeropoint offsets in the object spectra after the sky-line normalization has been applied. 
For each target F-star candidate, we fit a grid of \citet{Kurucz:1993uq} stellar models
to the stellar photometry in order to determine an appropriate
model.  
We convolve the best fitting stellar spectrum
and measure, through a $\chi^2$ search, any residual offset between the model and stellar wavelength model.  
Typically, the final shifts are small ($<0.1$ pixels) and vary smoothly as a function of position on the focal plane
(as one would expect if the shifts were due to alignment errors or large scale drifts in the astrometry).  
We fit the shifts as a function of position on the Baade focal plane and apply the zeropoint correction to 
each object spectrum.   

\subsection{Coadding Extracted Spectra}
Throughout the extraction and wavelength calibration process, we work on each individual science
exposure separately rather than on a coadded image of all of our exposures.  Primarily, we do this
to protect against guiding errors, which can accumulate from exposure to exposure, leading to 
systematic problems when trying to extract a coadded image.  Furthermore, as we work in 
nod \& shuffle mode, if a subset of the exposures were subject to changes in the sky brightness
on timescales shorter than our nod time, these images will have systematically poor sky subtraction;
including them before extracting will lead to systematic errors in our extracted spectra. 

Before coadding individual exposures, we first search for frames which do not have the quality
desired to be included in the final stacked science data.  If a frame was observed under conditions
in which the nod \& shuffle procedure fails due to sky conditions changing on timescales shorter than our nod
timescale, the sky subtraction will leave residual light (or deficient light)
in one object slit compared to the second.  This will manifest itself both in systematic errors
when the difference between the object spectra in slit A are compared to slit B, $O_{\rm A} - O_{\rm B}$.  Furthermore,
remnant sky light in the extracted spectra will lead to systematically different signal-to-noise
ratios in slit A, $SN_{\rm A}$, versus that in slit B, $SN_{\rm B}$.  We isolate any exposures
with significant difference from the mean in the $<O_{\rm A}-O_{\rm B}>$ versus $SN_{\rm A}/SN_{\rm B}$
plane and identify those exposures as possibly contaminated exposures and reject those
exposures when constructed the coadded final spectra.  Of the 705 science images we observed for PRIMUS,
12 exposures were rejected based on these criteria.

After any problematic exposures have been removed, we coadd the extracted one-dimensional spectra
 measured in each of our exposures to determine the final spectrum for each object.  
 In order to weight by the quality of the data in each exposure
(for example due to varying airmass or transparency between exposures), we weight each exposure
according to the mean signal-to-noise ratio of the mask. We construct this mean signal-to-noise ratio by
fitting the signal-to-noise of each object spectrum as a function of object magnitude and determine the value
of this trend at $i=21$.
We do not coadd the extracted spectra from slit A and slit B in each exposure.   If we were to coadd these
spectra, the fact that our objects are shifted both vertically and horizontally on the IMACS focal plane means we would
need to shift the two measurements spectrally in order to coadd and would result in the noise
in our final coadded spectrum being correlated.  While this correlation could be tracked and treated 
properly in the redshift fitting, we preferred to not coadd the two slits and simultaneously fit 
the spectrum from slit A and slit B
when solving for redshifts.   

\begin{figure}
\includegraphics[width=3in]{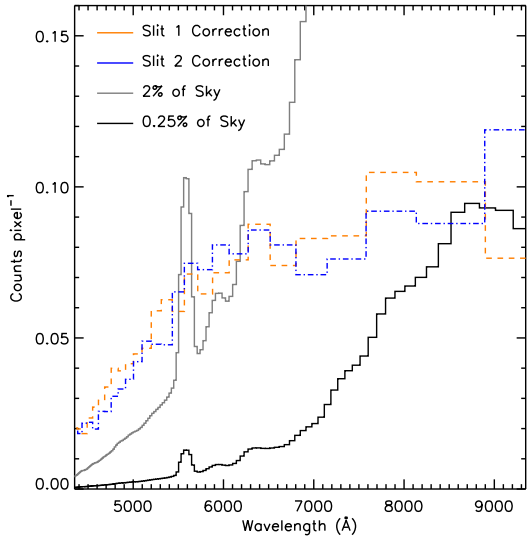}
\caption{\label{fig:emptycor} Magnitude of the residual light correction, derived by coadding 
the extracted spectra from slits in which no objects were observed, as a function of wavelength.  The orange (dashed) and blue (dot-dashed) lines show 
the median correction for slit 1 and slit 2 respectively.   The small scale correction contains about 1-2\% of the flux of the 
sky and thus is an important correction when studying faint objects.  }
\end{figure}

After our full extraction, scattered light corrections, and nod \& shuffle 
sky subtraction, we find that slits drilled without any objects in them show 
small-scale residual light.   In order to correct for this residual 
sky-subtraction error, we coadd the extracted counts in empty slits 
separately in slit A and slit B for each of the eight IMACS ccds for an 
entire night of observations.   This results in 16 wavelength-dependent 
corrections for each night (one for each of slit A and slit B on the 8 CCDs).  
We interpolate this mean residual to the wavelengths of each object in our 
coadded spectra and subtract it before proceeding with redshift fitting.   
Figure \ref{fig:emptycor} shows the magnitude of this effect compared to scaled
sky spectra.  Overall, this effect is 1-2$\%$ of the sky flux and thus an important
correction.
Removing this term greatly improves the final agreement between the 
extracted counts in slit A compared to slit B and redshift failures rates
for galaxies with known redshifts.
On nights in which, primarily due to 
hardware problems with the mask cutting, we had too few empty slits to make these measurements we utilized
the derived correction from the nearest night with viable data.

\section{Spectrophotometric Calibration}
\label{sec:1.5dcalibration}
Accurate spectrophotometry is essential for determining reliable
redshifts for PRIMUS because our spectral
resolution is too low to rely on resolve individual narrow emission lines;
therefore, for many objects the redshift is largely determined by the
spectral \emph{shape} (e.g., the strength of the $4000$-\AA\, break).
We emphasize that our objective is to obtain reasonably
accurate \emph{relative} spectrophotometry; the \emph{absolute}
normalization of each spectrum, which depends on the amount of light
lost from the slit due to the physical extent of each object relative
to the slit width, variations in the PSF, pointing errors, and so
forth, is not relevant for determining the redshift with our method.

Traditionally, standard-star observations are used to derive a
sensitivity function to flux-calibrate
the observed spectra.  By contrast, we approach the problem of
spectrophotometric calibration from a forward modeling
perspective.  In other words, we apply our best estimate of the
wavelength-dependent throughput to the \emph{model} templates used for
redshift fitting (see \S\ref{sec:1dfitting}).  Thus our principal goal
is to determine the throughput as a function of both position on each CCD
and time; we then employ the throughput estimate when fitting for
redshifts.

\begin{figure}
\includegraphics[width=3in]{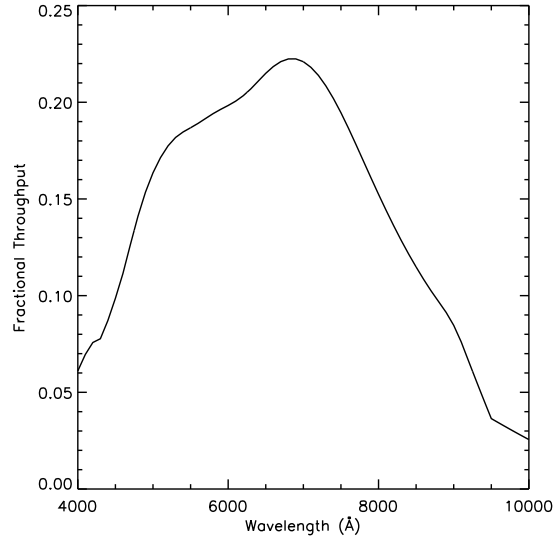}
\caption{\label{fig:imacsthru} Baseline IMACS throughput used as a zeroth-order flux calibration term throughout the PRIMUS flux calibration.  This curve was obtained
by observing a DA white dwarf on CCD 5 of the IMACS array. The broad absorption line features allowed us to perform wavelength calibration on the slitless spectrum.  
We utilize the observed sky emission in each PRIMUS slit to extend the flux solution between each of the IMACS CCDs and tie each CCD to the same flux solution.  We finally
apply a correction based on the $\sim30$ F-stars we observe on each mask to remove any variations from the overall IMACS response function over the course of the survey's operations.}
\end{figure}

We first utilized slitless DA white dwarf observations taken
in Jan 2006
to obtain an overall throughput curve for the IMACS
instrument and PRIMUS prism as shown in Figure \ref{fig:imacsthru}.
We are able to calibrate the
wavelength of the extracted slitless spectrum from the broad Balmer
lines present in the DA spectrum. Next, we 
correct the spectra for atmospheric extinction at the mean airmass of
each mask utilizing the Gemini South atmospheric extinction curve.
Clearly one correction curve 
obtained at the beginning of the survey is not appropriate to fully calibrate
our data in subsequent runs; we utilize this normalization as a baseline.  More detailed
corrections are applied to the data based on observations of F-stars and the night sky to 
remove the higher order fluxing residuals from this baseline correction.

In order to quantify how the flux correction varies between each of
our PRIMUS observing runs, we utilize two tracers for the sensitivity
measured on each of our PRIMUS masks.  We utilize the $\sim30$ F-stars
on each mask to measure an average overall sensitivity function for
each PRIMUS run and use the sky spectra in each mask to correct for
any variations in the inter-ccd sensitivity corrections.  While, in
practice, we could generate mask by mask corrections, the variations
between masks in a given run are small;  averaged corrections allow the
use of more F-star and sky measurements when constructing our fluxing terms leading
to higher signal-to-noise ratio corrections.

\begin{figure}[b]
\includegraphics[width=3.5in]{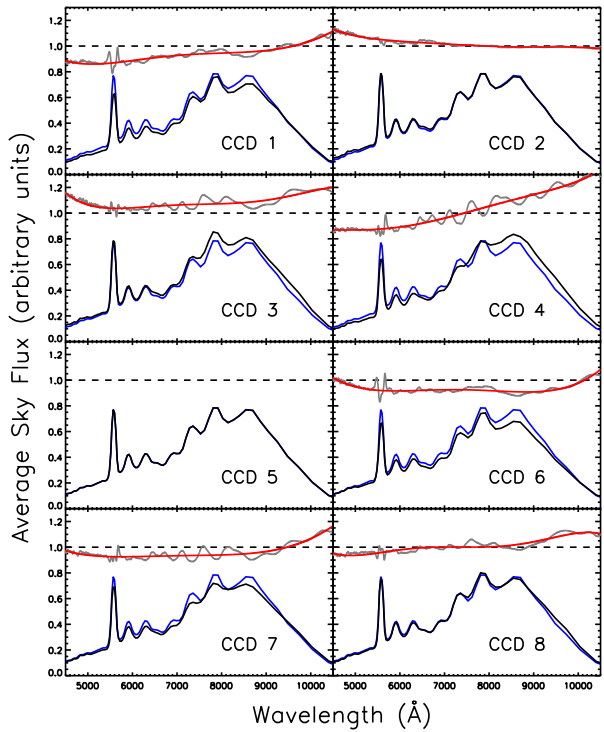}
\caption{\label{fig:skyfluxmask} Derivation of the inter-ccd spectrophotometric corrections for one PRIMUS mask.  The baseline IMACS 
throughput and comparison to F-stars 
observed on PRIMUS masks ties the PRIMUS spectra spectrophotometrically to IMACS CCD 5.  Based on observations of the sky spectrum 
while in the ``off'' position in our nod \& shuffle observations, we correct for the response differences between CCDs by constructing 
the mean sky spectrum on each CCD (blue). The ratio between this mean spectrum and the mean sky spectrum from slits on CCD 5 (black) 
gives a spectrophotometric correction to bring each ccd to the same system (grey). Lastly, we fit the observed ratio with a smooth 
bspline to characterize the low-order response variations between CCDs (red).  As the temporal variation of these corrections is low, we combine observations over
each month of PRIMUS observations when calculating the final inter-ccd flux correction to provide the highest signal-to-noise ratio correction possible.}
\end{figure}

\begin{figure}
\includegraphics[width=3.5in]{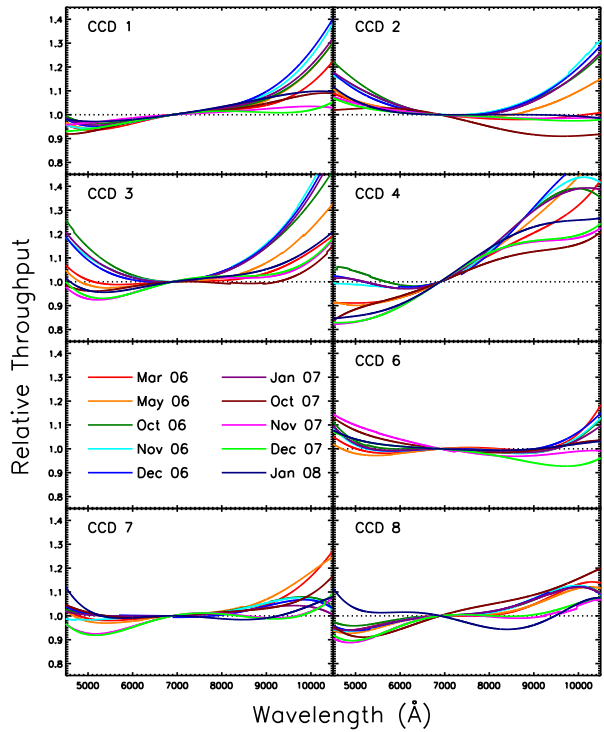}
\caption{\label{fig:skyfluxcompare} Variation of the inter-ccd flux corrections described in the text and illustrated in Figure \ref{fig:skyfluxmask} as a function of time.
For each CCD, we show the final mean correction used to correct for variations between CCDs for each month of PRIMUS observations.  Overall, the general form of the correction
remained constant throughout the survey, but modest changes occur primarily due to instrument maintenance throughout the course of the survey.}
\end{figure}

The baseline IMACS sensitivity is tied to CCD 5 in the IMACS array.  In order to normalize each of the other CCDs, we utilize the sky spectrum observed in the 
``off'' position in our nod \& shuffle observations.  Based on the assumption that the sky spectrum is constant over the IMACS field of view,
we measure the mean sky spectrum on each of the eight IMACS CCDs.  We then measure the ratio between the mean sky measured on CCD 5 to the mean sky measured
on each of the other CCDs.  Figure \ref{fig:skyfluxmask} shows this procedure for one PRIMUS mask. The black spectra
show the CCD 5 mean sky spectrum and the blue shows the mean for each of the other CCDs.  The red line is a bspline fit utilized to remove the 
variation between IMACS ccds. Note that we assume the sky spectrum does not vary over the IMACS field of view.  While this is not an 
unreasonable assumption for the sky continuum, emission line strengths can vary on arcminute scales. The grey lines in Figure \ref{fig:skyfluxmask}
illustrates this issue. As we ultimately fit a smooth bspline to the data, the few percent variations in the sky emission lines 
do not significantly affect our methodology. We do not expect the relative normalizations
between the CCDs to be a strong function of time; by combining data over multiple masks and nights allows us to construct the highest possible signal-to-noise
ratio correction.  After examining the time dependence of these corrections, we chose to create these ``inter-ccd" corrections for each month of PRIMUS observations.
Figure \ref{fig:skyfluxcompare} shows the variation with time of these correction terms.  Overall, the curves for each CCD follow the same general shapes over the
full course of the PRIMUS observations. Some differences arise likely due to modifications to the instrument and electronics over the course of our observations.

\begin{figure}
\includegraphics[width=3.5in]{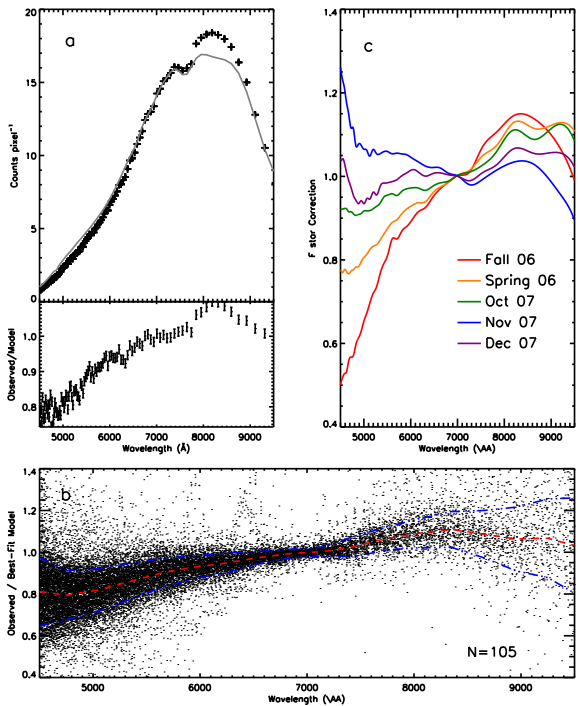}
\caption{\label{fig:fstarflux} Derivation of overall flux calibration with PRIMUS F-stars. Panel a: The ratio of a PRIMUS-observed F-star 
(black datapoints) and model F-star spectrum convolved to PRIMUS resolution (grey) is used to create the correction needed to bring PRIMUS 
spectra to a standard system.  Panel b: To obtain the highest signal-to-noise ratio measurement of the flux calibration, we coadd the 
individual corrections over PRIMUS observing seasons to create the final mean correction.  The raw
data show the measured ratio between observed spectrum and model \citet{Kurucz:1993uq} spectrum.  The final mean correction (red dashed line) 
is obtained by finding the mean of the individual corrections. The blue dot-dot-dot-dashed line shows the $1\sigma$ region around the mean. 
 Panel c: The temporal variation of the F-star flux calibration vectors.  Throughout the course of the survey, the IMACS team made several 
improvements and performed regular maintenance to the instrument -- the results are clear in the increase blue sensitivity of the IMACS CCDs throughout the course of the survey.}
\end{figure}

With each of the CCDs on the same spectrophotometric system, we next bring the total spectrophotometric system of each mask onto the final
survey system.  While the baseline IMACS correction corrects for a general response function of the instrument, this response function can vary over time.  
In order to remove this variation, we utilize the $\sim30$ F-stars we observed on each mask as our spectrophotometric standards.
For each F-star, we fit its broadband photometry against a model
of \citet{Kurucz:1993uq} stellar atmosphere models.  We measure the ratio between
the observed PRIMUS spectrum and the best-fitting stellar model which has been 
convolved to PRIMUS resolution.   Figure \ref{fig:fstarflux} (a) shows this fitting process;
the datapoints with errorbars show the PRIMUS data and the grey line is the convolved \citet{Kurucz:1993uq}
model.  The resulting correction vector is the ratio between the observed and model data.  For each PRIMUS
observing season, we combine each of these correction vectors to derive a median correction vector for the entire
run as shown in Figure \ref{fig:fstarflux} (b).  After careful comparisons, we found that the fluxing corrections
within a single observing season were consistent, but corrections between seasons shows considerable
variation.  Figure \ref{fig:fstarflux} (c) shows the resulting flux corrections for each of the PRIMUS observing seasons.  
The changes illustrated in the figure are often physically motivated. Several major modifications were completed to the IMACS
instrument throughout the course of our observing program 
to increase the throughput and sensitivity of the instrument.  The corrections
clearly shows that these adjustments made sizable improvements, especially at blue wavelengths.

All of these corrections correct the spectral shape of the PRIMUS spectra.  The main source of small-time variation in the flux correction of our spectra
which is not removed in this analysis is due to transparency variations throughout each night.  These transparency variations 
will drive the overall normalization of our spectra high or low but will not affect the spectral shape of the spectra.  
When fitting our redshifts, as described in \S\ref{sec:zfitting}, we tie the normalization of our final model to the 
broadband photometry and allow the normalization of the spectra to be set by the photometric data, so transparency variations 
do not affect the quality of our redshift fits.

% ####################
\section{Determining Redshifts}
\label{sec:1dfitting}
\subsection{Galaxy Spectral Templates}
When fitting PRIMUS spectra, we chose to compare
each PRIMUS spectrum to an empirical library of galaxy spectra in order to measure the 
best fitting galaxy redshift for each object.  We use empirical templates, rather than a more
general non-negative linear combination of spectral models.  At PRIMUS resolution,
fine spectral features cannot be utilized to help break degeneracies between two models which may produce
similar low-resolution spectra but significantly different when viewed at higher resolution.  
Thus, relying on empirical galaxy templates allows us to fit our PRIMUS sample 
while  preventing the best
fitting models from occupying areas of parameters space which have no known counterparts in reality.  

In order to build our empirical library of galaxy templates, we use the AGN and Galaxy Evolution
Survey (AGES) galaxy sample \citep{Kochanek:2011}.  AGES observed
18163 galaxies with a median redshift of $z\sim0.3$  in the 
NOAO Deep Wide-Field Survey \bootes field.  
This sample is an ideal source for our empirical templates due to the high quality spectroscopy to $I=20.4$ and at higher redshifts than
probed by lower-redshift surveys such as SDSS.  

When constructing the sample of AGES galaxies to be considered in our library building, 
we limit the sample to galaxies with $0.05<z<0.35$ and only keep spectra with a mean
signal-to-noise ratio of at least 4 per pixel in the AGES spectrum. This redshift range ensures
that H$\alpha$ fell in the observed AGES wavelength range.  Furthermore, to
ensure high-quality spectrophotometry and to create a statistically 
complete sample, we only consider objects with $14.5<I_{\rm Vega}<19.5$.  
We also only included AGES observations that had robust spectrophotometry.
These quality cuts
leave a sample of 3244 galaxies which will refer to as the ``AGES parent" sample.

As the observed spectra from AGES do not extend sufficiently into the blue to 
cover the full observed PRIMUS wavelength range when shifted to higher redshifts, we
instead use a best-fitting spectral model for the parent sample galaxies when
deriving the empirical library.
For each galaxy in the AGES parent sample, we perform a detailed spectral 
modeling of the AGES spectra simultaneously with the measured optical, GALEX, and
{\it Spitzer} photometry.  We utilize the continuum and emission line fits 
to AGES galaxies from \citet{Moustakas:2011} 
when constructing the PRIMUS galaxy template set; the full details of the methods involved in the
fitting of the AGES spectra can be found in \citet{Moustakas:2011}.  
The  best
fitting parameters were then used to construct a full UV to IR model for each AGES galaxy which
is used throughout the rest of this analysis.  

In order to construct a library of empirical templates from the best fitting models, we create
a subsample (``basis set" hereafter) of the full AGES sample which, when convolved to the PRIMUS resolution, fully
span the parent sample.  To construct this sample, we first convolve each AGES sample galaxy to PRIMUS resolution
assuming it was observed at $z=0.25$.  We next compute a 3244x3244 matrix from these convolved spectra; each element, 
$c_ij$, in this matrix is the $\chi^2$ value calculated assuming galaxy $i$ could be described by a simple scale factor 
multiplication of galaxy $j$ in the AGES sample.  One pair of spectra can represent each other in our
basis set approach if the $\chi^2$ is smaller than a given maximum tolerance $\chi^2_{\mbox{max}}$.  Based on this
matrix, we strive to determine the minimal number of AGES sample galaxies that will span the entire AGES population. 

When constructing the PRIMUS basis set, we set $\chi^2_{\mbox{max}}=2$.  We then set each element in a new 3244x3244 matrix
to 1 if $\chi^2_{i,j}<\chi^2_{\mbox{max}}$ and 0 if the calculated $\chi^2$ was larger than our threshold value.  We then minimize
the number of columns subject to the condition that each row must have at least one element larger than 0.  Mathematically,
we solve
\begin{equation}
	\mbox{min} \sum_j x_j
\end{equation}
subject to 
\begin{equation}
	\sum_j \alpha_{ij} x_j \ge 1
\end{equation}
where $x_j=1$ if column j is a basis galaxy, $x_j=0$ otherwise, and $\alpha_{ij}$ is the element in our binary matrix.
This method results in 206 basis set galaxies. 

\begin{figure}
\includegraphics[width=3.5in]{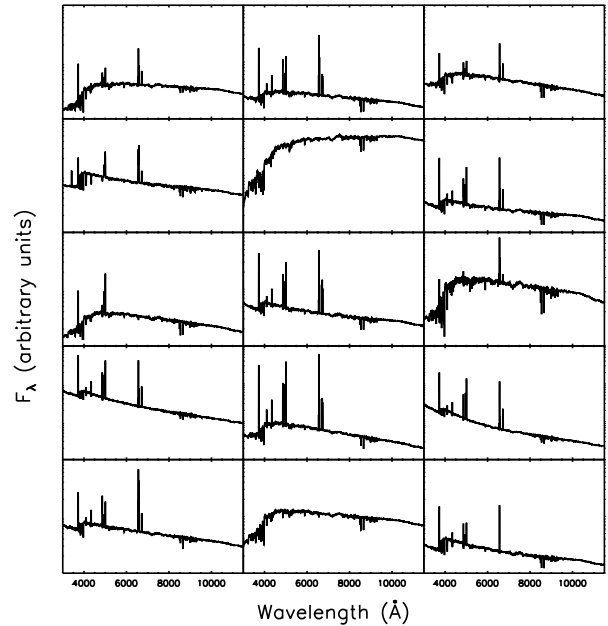}
\caption{\label{fig:basisexample} Random selection of 15 basis galaxies from the template used to fit PRIMUS spectra.  Basis galaxies were selected by asking for the set
of galaxies from fits to AGES objects which fully spanned the AGES galaxy sample if convolved to PRIMUS resolution. 
  As the majority of the diversity in galaxy properties arises from emission 
line strengths and continuum shapes in blue galaxies, most of the PRIMUS basis galaxies are star-forming systems.}
\end{figure}

While the galaxy models utilized to construct this basis set is complete in galaxy properties over the range of properties probed 
by the parent sample of AGES galaxies, the redshift limitation when constructing the AGES parent removes any information 
about the 2700$\AA$ MgII emission line strengths from our basis set.  To remedy this, we isolated any basis set galaxies with 
$^{0.1}(u-g)<1.0$ and $EW_{[OII]} <15$  to be possible MgII emitters based on AGES galaxies that have 
detected MgII emission at $0.4<z<0.9$ and added 
a second basis galaxy to the set identical to the first with the exception of the addition of a 25 $\AA$ equivalent 
width MgII line. This duplication is valid as we are simply seeking a set of galaxies which span the properties of galaxies 
at PRIMUS resolution but we do not require the minimal set of such galaxies. Figure \ref{fig:basisexample} shows 15 randomly 
selected galaxies from the basis set used when fitting PRIMUS galaxies.  As the largest source of spectral diversity in the 
galaxies, once convolved to PRIMUS resolution, is due to the relative strengths of optical emission lines, the basis set is 
primarily composed of blue star forming objects, while the red galaxy population is well sampled with only a small number of basis objects.

\begin{figure}[b]
\epsscale{1}\plottwo{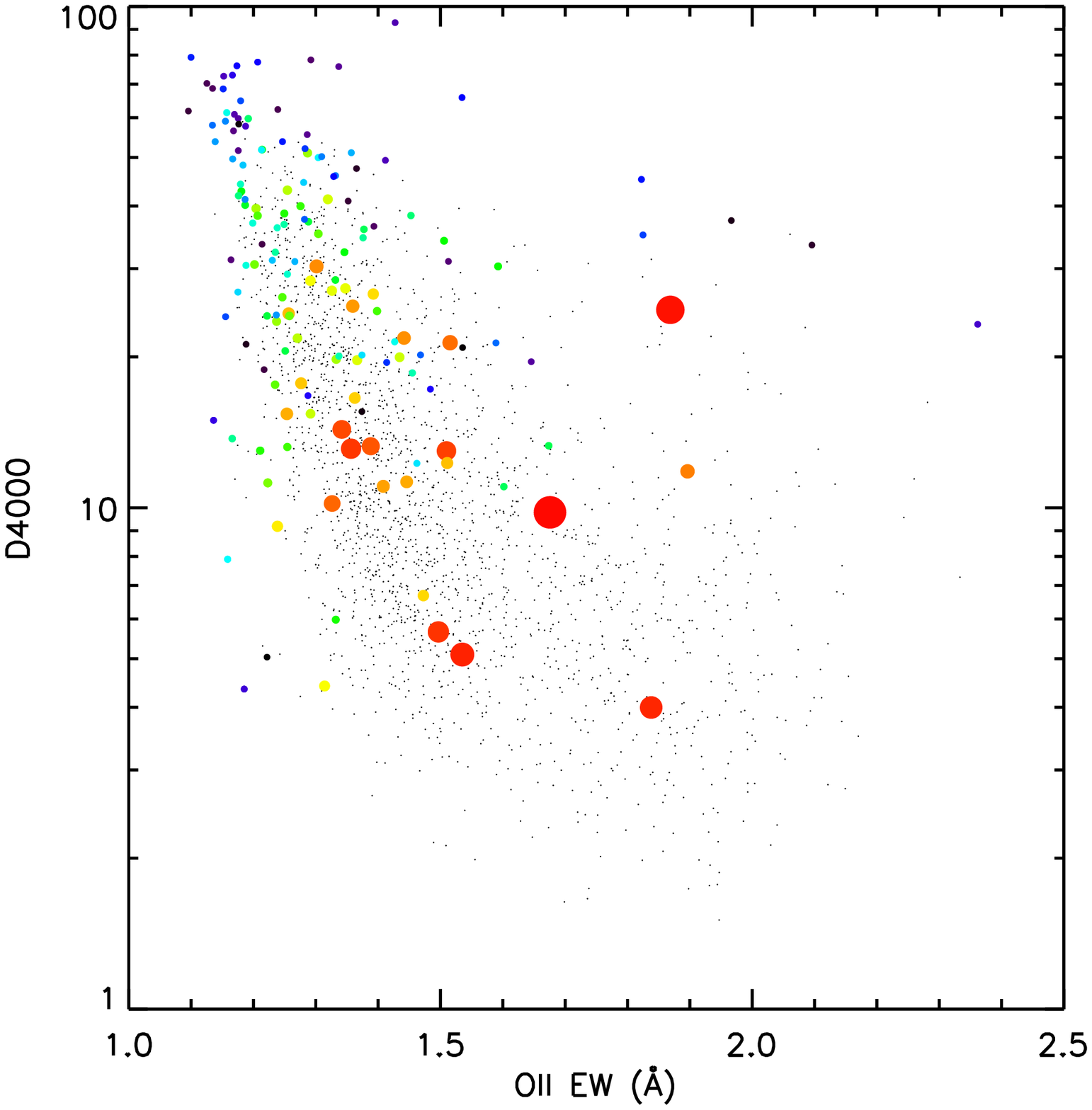}{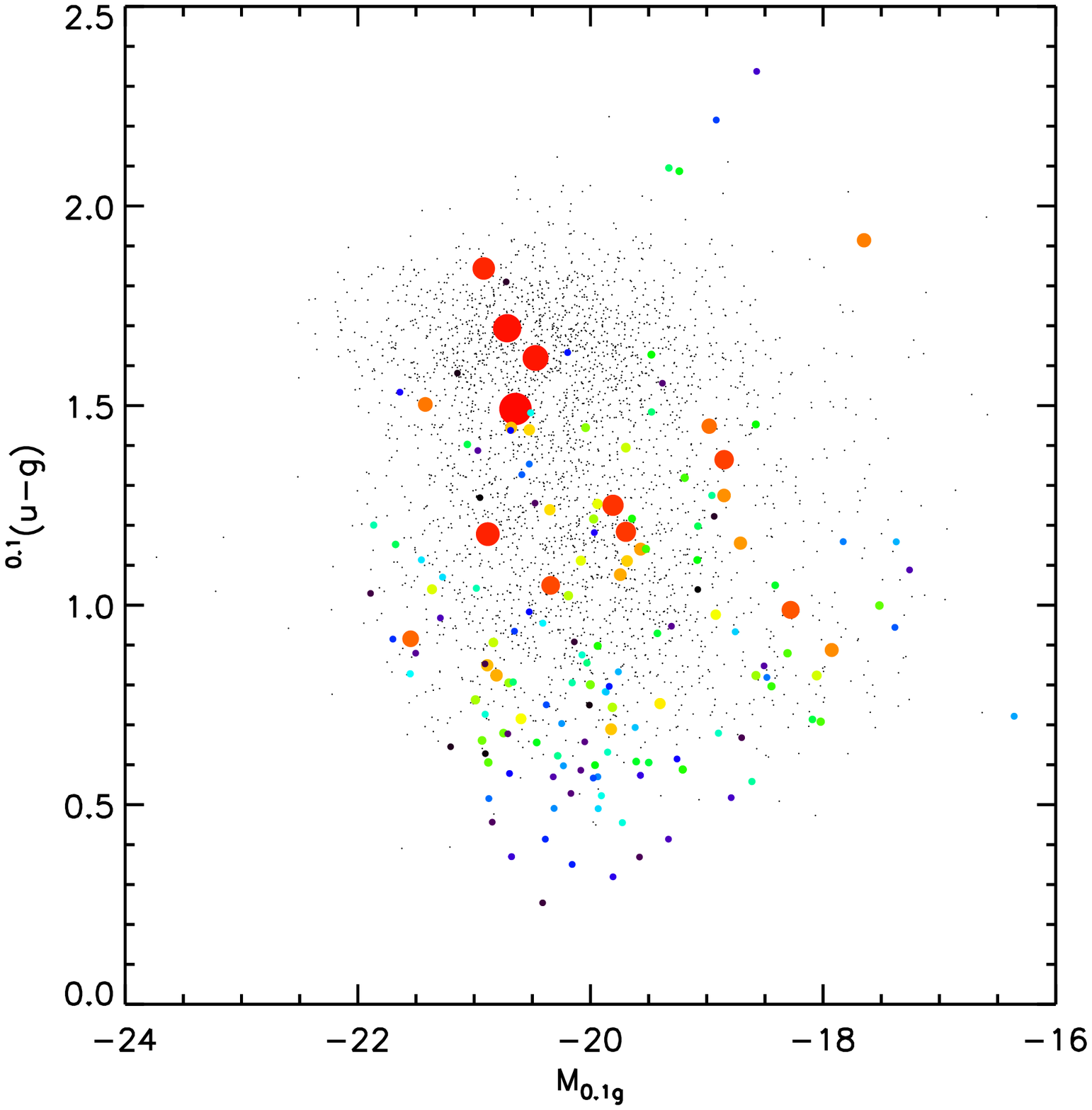}
\caption{\label{fig:basisprop} Properties of the PRIMUS basis sample (colored points) compared to the space spanned by the full AGES 
galaxy sample.  In both panels, the color and size of data points shows the basis weight for each basis galaxy.  Redder (larger) data
 points represent basis galaxies that can be used to represent more AGES sample galaxies when convolved to PRIMUS resolution.  The vast
 majority of the variation in this basis set is in the blue/star forming galaxies.  Left: Comparison of the strength of the 4000 $\AA$
 break, D4000, with the equivalent width of \oii.  Right: Coverage of the PRIMUS basis galaxies in restframe color-magnitude space.}
\end{figure}

While our final basis set spans the full range
of galaxy types in AGES when observed at PRIMUS resolution, 
some galaxies in this basis are much more common than others.  In order to quantify this
for later use in the PRIMUS redshift fitting, we utilize the basis set to fit the full AGES parent sample.
For each basis galaxy, we count the number of parent galaxies that were best fit by it and record this value as
the basis weight of the template, $b_i$.  Figure \ref{fig:basisprop} shows the distribution of the full AGES parent sample (small black points) and
the PRIMUS basis set. The size and color of the symbols represent the basis weight of each template (redder/larger symbols show basis 
galaxies which characterize more parent sample 
galaxies at PRIMUS resolution). In both the D4000 versus \oii EW and color-magnitude parameter space, 
basis galaxies clearly dominate the blue/star forming portions of the diagrams; these galaxies show a large diversity in 
emission line strengths and continuum shapes compared to more quiescent galaxies.  The red/quiescent basis galaxies, on 
the other hand, have very high basis weights; these basis objects describe a large fraction of the AGES parent set when convolved to PRIMUS resolution.

\subsection{AGN and Stellar Spectral Templates}
In addition to galaxies, PRIMUS observed broadline
AGNs; Paper I includes the details of the AGN selection in PRIMUS, but in brief
we often did nothing to remove AGN from our galaxy sample and in some fields we actively targeted objects with color
characteristic of AGNs.  In order
to measure redshifts for these objects and ensure that only galaxies are being assigned
redshifts from galactic templates, we also fit each PRIMUS object with broadline AGN templates during redshift fitting.

Obscured AGNs in which the optical spectrum has contributions from narrow emission lines from the 
AGN but is not dominated by broad-line emission are fit using the galaxy templates.  The AGES parent
sample contains examples of these Type II AGN and the stellar continuum dominates the broad shape of the
spectra of these objects, and thus the galaxy templates are sufficient for determining redshifts.  It 
should be noted, however, that due to the low resolution of PRIMUS, the presence of narrow emission lines
from AGNs does not allow us to classify galaxies as AGNs. Throughout this section, we focus on 
creating a set of broad-line AGN templates used to classify objects which are dominated by the
non-stellar AGN continuum and broad emission lines.  

As a basis for our set of AGN templates, we use the SDSS composite quasar spectrum from \citet{vdb2001_composite}.
While this composite contains the broad average of quasar features, the full population of AGNs have 
a wide range of spectral indices, $\alpha$, and reddening due to dust.  To span this
range of parameters, we create a grid of AGN models based upon the \citet{vdb2001_composite} composite but
for which we have modified the spectral index of the AGN continuum to values of $\alpha=-2.5, -2.0, -1.5, -1.0, -0.5, 0.0$ 
and have included dust reddening with $E(B-V)=0.0, 0.05, 0.1$ assuming a dust model similar to the 
SMC. The range of $E(B-V)$ values agrees well with the observed range of reddening observed in SDSS quasars in 
\citet{Hopkins:2004}.
  We find that the population of AGN in the PRIMUS sample is well spanned by these modifications in $\alpha$ and $E(B-V)$ and tests with 
broader ranges in each parameter showed the majority of PRIMUS AGN to be best fit inside the final range specified above.

In addition to the variety of continuum shapes found in AGNs, as one observes AGNs at increasingly higher 
redshifts, attenuation from the intergalactic medium (IGM) plays a critical role in shaping the rest-frame UV (observed optical) spectra.
  Neglecting this attenuation can lead to large differences between model spectra and observed
AGNs at $z>3$.  With the depth probed by PRIMUS, we expect many quasars at these redshifts to be present in the sample,
and thus we include a redshift-dependent attenuation from the IGM in our library of AGN template spectra based on an 
the redshift evolution of the IGM attenuation curve presented in \citet{Madau:1995}.

In addition to galaxies and AGN, the straightforward flux limit cuts in the  PRIMUS target selection include
galactic stellar sources as well, especially at faint magnitudes where photometric separation between stars and
galaxies becomes more difficult.   We utilize the 
empirical \citet{Pickles1998_stellarlib} stellar library to create a sample of stellar 
templates which we compare to each observed PRIMUS spectrum in order to determine if the source is stellar in nature.  
Due to the low resolution of PRIMUS spectra, subtle changes in the 
stellar absorption features or spectral shape are not important for our application; the goal with these
stellar spectra is to ensure that stellar objects are not misclassified as galaxies when we fit redshifts.

\begin{figure}
\includegraphics[width=3.5in]{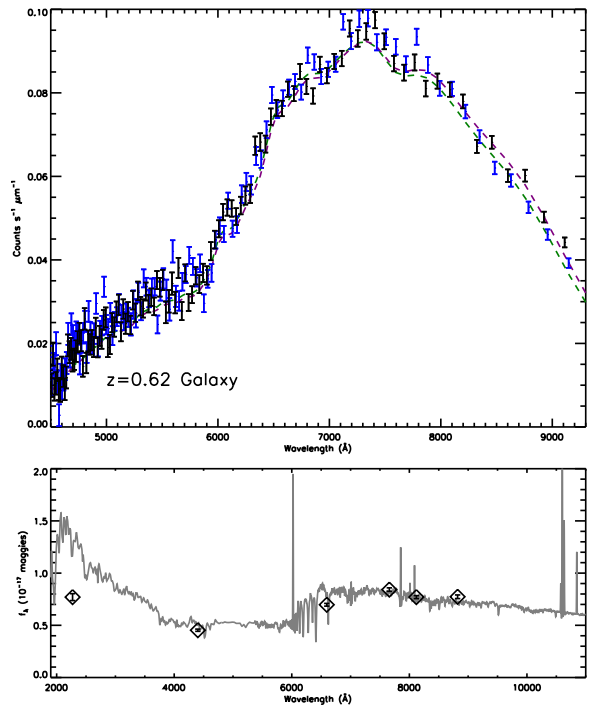}
\caption{\label{fig:fitgalaxy} Example of a PRIMUS fit to an object identified as a $z=0.62$ 
galaxy. The top panel shows the observed PRIMUS spectrum from both slit A (black) and slit B (blue). 
The dashed line shows the best-fitting model convolved to PRIMUS resolution.  In the bottom panel,
the grey line shows the full-resolution best-fitting galaxy model. The data points with errorbars mark 
the broadband UV and optical photometry available for the object.}
\end{figure}

\subsection{Fitting Method}
\label{sec:zfitting}
With the spectral template sets constructed, we perform redshift fitting of
each PRIMUS object.   Throughout the fitting process, we fit the slit A
and slit B observations from our nod \& shuffle method simultaneously.  In addition to the PRIMUS
spectra, we also simultaneously fit the observed optical photometry of each source. 
The addition of the photometry both adds information outside the spectral window
of the PRIMUS spectroscopy and allows us to find the best fit in the possible
presence of long-wavelength spectrophotometry errors.  
As we have not corrected for the large (and highly uncertain) slit losses in our PRIMUS
spectra, we do not force the simultaneous fitting of the photometry and spectra to agree
in absolute normalization.

When fitting galaxies, we first fit each object on a coarse grid with $\Delta z=0.01$ spacing
from $z=0$ to $z=1.2$.  AGN are fit from $z=0.0$ to $z=5.0$ with 181 grid points such that the $i$th 
grid point, $z_{{\rm AGN}, i}$, is 
\begin{equation}
	1+z_{{\rm AGN}, i} = 1.01^i.
\end{equation}
At each grid point, we calculate the $\chi^2$ between the measured photometry (with associated errors) and
the PRIMUS spectra.  When fitting, we do not give the photometry more weight than is given by the associated errors; in 
essence, photometry is adding $\sim$6 more data points of information to the 300 data points available from spectroscopy and thus does not dominate the fit.  Due to possibility of long-wavelength spectrophotometry errors in the PRIMUS 
spectra, we fit each object with a model of the form
\begin{equation}
	M(\lambda) = a_0M_0(\lambda)\left(a_1\xi(\lambda)+a_2\xi^2(\lambda)\right)
\end{equation}
where $M_0(\lambda)$ is the original template spectrum and $\xi(\lambda)$ is a 
refluxing vector that account for possible errors
in the spectrophotometry. In detail, $\xi(\lambda)$, is a term that characterizes
a linear term in ${\rm log} \lambda$
\begin{equation}
	1+\xi(\lambda) = 2\frac{{\rm log}\left(\lambda/\lambda_{\rm min}\right)}{{\rm log}\left(\lambda_{\rm max}/\lambda_{\rm min}\right)}.
\end{equation}
When performing the fits, we place priors of 10\% on $a_1$ and $a_2$; while we expect some level of 
spectrophotometric errors on large-wavelength scales, we do not want the fitting to be dominated by
driving the refluxing terms to unreasonable levels to compensate for actual differences between the selected
model and the observed spectrum. This refluxing only applies to the comparisons between the PRIMUS spectra
and convolved model.  The photometry are not fitted with any form of refluxing.

Based on these fits
we calculate the $\chi^2$ between each template spectrum and the observed photometry and spectroscopy.  
For stars, we choose the best stellar fit to be the stellar model with the lowest $\chi^2$.  For AGN, 
we choose the best fitting AGN template at each grid point by choosing the lowest $\chi^2$ model.  For galaxies,
as we have information about the real frequency of each galaxy template in the full AGES parent sample. Rather
than choosing the lowest $\chi^2$ model, denoted $\chi^2_{\rm min}$, at each grid point, we calculate the effective $\chi^2$, $\chi^2_{\rm eff}$, 
based on the basis weights, $b_i$
\begin{equation}
	\chi^2_{\rm eff} = \chi^2_{\rm min} - 2\, {\rm log} \left( \sum_j b_j \,{\rm exp}\left(-\frac{\chi^2_j-\chi^2_{\rm min}}{2}\right)\right).
	\end{equation}

For galaxies and AGN, we find the minimum in $\chi^2$ versus redshift in our grid.  We then fit a finer linear
grid with $\delta z=0.001$ in an interval of $|z-z_{\rm min}|<0.03$ around the minimum found in the course grid, 
$z_{\rm min}$. Based on this finer grid, we calculate the final redshift of the galaxy or AGN as 
\begin{equation}
	z_{\rm best} = \int z\,{\rm exp}\left(-\frac{\chi(z)^2-\chi^2_{\rm min}}{2}\right)dz
\end{equation}
and the associated error is the second moment of the probability distribution
\begin{equation}
   \sigma_z = \int z^2 	\,{\rm exp}\left(-\frac{\chi(z)^2-\chi^2_{\rm min}}{2}\right)dz.
\end{equation}

Once we have fit each PRIMUS object with galaxy, AGN, and stellar models, we
must choose which of these models is the best representation of the measured data. 
If the final star or galaxy  $\chi^2$ is the lowest among the models, we choose
to classify the PRIMUS object as a star or galaxy.  After manually inspecting 
objects with $\chi^2_{\rm AGN} < \chi^2_{\rm gal}$, including objects with known
redshifts and classifications from DEEP2 and zCOSMOS, we find that
a threshold of $\chi^2_{\rm AGN} + 100 < \chi^2_{\rm gal}$ provides correct redshift identification for objects that 
are clearly AGN (either from inspection of PRIMUS spectroscopy or previous high-resolution spectroscopy).

\begin{figure}[b]
\includegraphics[width=3.5in]{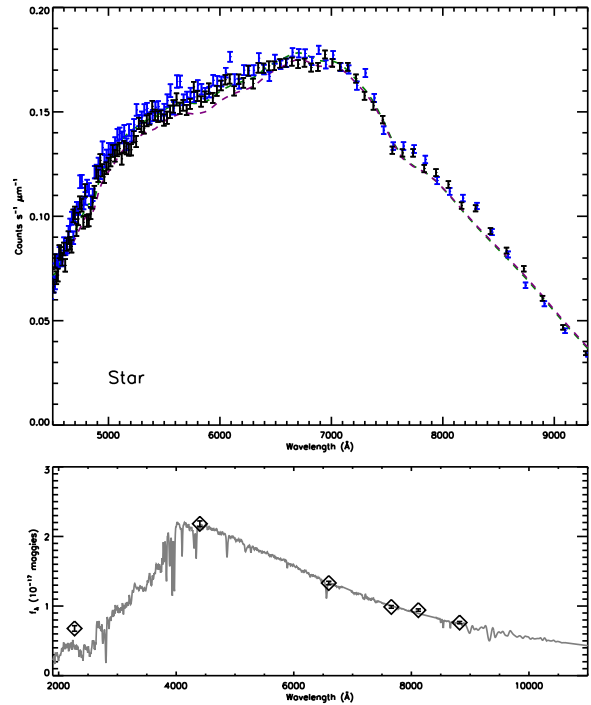}
\caption{\label{fig:fitstar} Same as Figure \ref{fig:fitgalaxy}, but an example of a PRIMUS fit to an object identified as a star.}
\end{figure}

\begin{figure}
\includegraphics[width=3.5in]{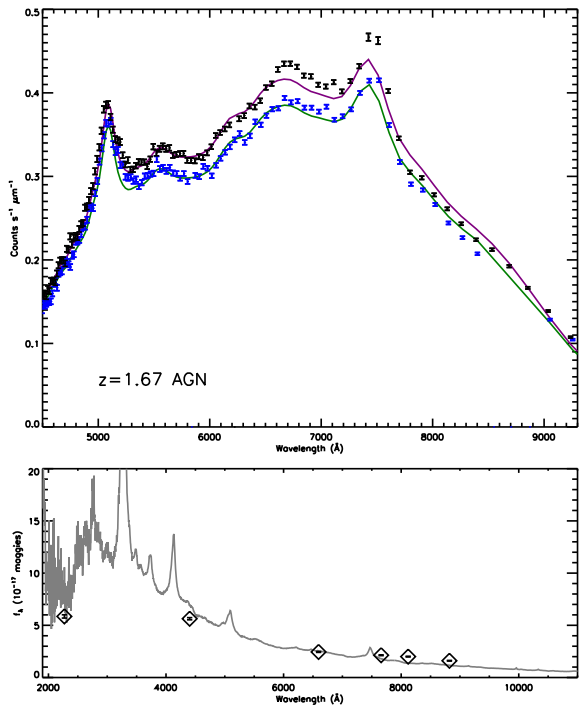}
\caption{\label{fig:fitagn} Same as Figure \ref{fig:fitgalaxy}, but an example of a PRIMUS fit to a $z=1.67$ AGN.}
\end{figure}

Figures \ref{fig:fitgalaxy}, \ref{fig:fitstar}, and \ref{fig:fitagn} show example PRIMUS fits. 
In each plot, the top panel
shows the PRIMUS spectrum observed through slit A (black) and slit B (blue) as well as the best-fit
spectrum convolved to PRIMUS resolution (dashed line).  The data bars on the PRIMUS spectra
are counting statistics and do not include possible systematics from sources such as
flat fielding or errors in our scattered light model. The bottom panel shows the best-fitting
high-resolution model as well as the broadband photometric data points available for each object (data
points with errorbars).

% ####################
\section{Redshift Accuracy}
\label{sec:redshiftaccuracy}

\subsection{Quality Control}
\label{sec:quality}
Before using the PRIMUS redshifts for scientific work, it is vital
to remove objects which could be contaminated due to systematics from the
extraction or redshift fitting.   

The majority (90\% see Paper I) of PRIMUS
redshifts do not change when photometry is included in the fit, however 
the addition of photometric measurements to the PRIMUS
fits does improve the spectroscopic fits of the remaining 10\%, especially at fainter magnitudes.  We thus require
that any robust redshift measurement from PRIMUS include at least three bands of photometry.
While some of the redshifts for objects with fewer than three bands of photometry may be correct, we find the failure rate
for objects with less than three bands of photometry is dramatically higher than for 
objects with more bands.  Similarly, we flag any objects with more than 70\% of the fitted
wavelength range masked in the PRIMUS spectroscopy.  These objects are predominately objects
that fell near the edges of CCDs on the IMACS focal plane.  Clearly, missing large sections of the
object spectrum limits our ability to measure a robust redshift for these objects.

We further flag objects if the fitting of the spectrum shows signs of possible problems.   If the best
fitting redshift was at the extremes of the redshift range fit $z=1.2$ for galaxies or $z=5.0$ for AGN the 
object is flagged as a possible problem. If the best fitting redshift is at the extreme of the fitting
range then we are unable to robustly measure a maximum in the probability distribution function. Additionally, 
we flag galaxies if the best fitting galaxy redshift is less than a $5\sigma$
($\Delta \chi^2<25$) detection compared to other redshift minima in the grid.  Typically, these are
low signal-to-noise objects with very flat probability distribution functions in redshift and the 
minimum is a noise spike. Lastly, we flag any objects with $\chi^2>10^4$; these are objects for which
no model in our library performed well in fitting the object and thus no redshift information 
from these objects is trusted.  

The final test we perform checks the total available information available in the measured photometry and
spectroscopy.   We fit each object with a grid of powerlaw models with indices running from $\alpha=-3.0$ to 
$\alpha=3.0$.  If our spectra and photometry provide little information beyond a
simple powerlaw model with no spectral features, we do not trust the measured redshift for the object.  If the best
fitting powerlaw has a $\chi^2$ of $\chi^2_{\rm best}-\chi^2_{\rm powerlaw} > -10$ (i.e the best fitting 
template model isn't better than $\sim 3\sigma$ compared to a simple powerlaw) we flag the object as suspect and
remove it from our final statistical sample. 

\begin{figure}
\includegraphics[width=3.5in]{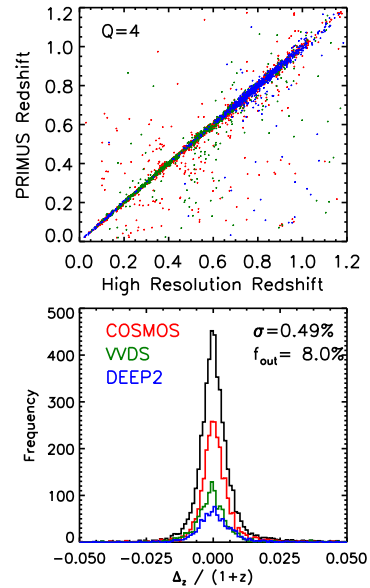}
\caption{\label{fig:zvzq4} Comparison of PRIMUS redshift with redshifts obtained from higher-resolution spectroscopy
from COSMOS, DEEP2, and VVDS for $Q=4$ quality rating.   The bottom panel shows the histogram of the difference between
the PRIMUS and high-resolution redshift for each survey as well as the full comparison sample.  Overall, we find a $\sim0.5\%$
dispersion between PRIMUS and the high-resolution sample with $\sim8\%$ of the galaxies falling more than $5\sigma$ from agreement.}
\end{figure}

Once we have a clean sample of PRIMUS galaxies we assign each object a numerical redshift quality value.  For each
galaxy, we construct the statistic,  $\zeta$, which quantifies the separation of multiple minima in redshift
space for each galaxy and AGN in the sample.  Explicitly, we measure the difference in depths of the best and 
second minima in $\chi^2$ space, $\Delta\chi^2$
\begin{equation}
	\Delta\chi^2 = \chi^2_{\rm 2nd} - \chi^2_{\rm best}
\end{equation}
and calculate 
\begin{equation}
	\zeta = 1000 \frac{\left(\sigma_z/(1+z_{\rm best})\right)}{\left(\Delta\chi^2\right)^{1/2}}.
\end{equation}
The numerator quantifies the width of the best peak in the $P(z)$ distribution and the denominator 
quantifies the separation, in a statistical sense, between the goodness of fit in the first and second minima.  
In extremes, if an objects has a very broad peak in $P(z)$ or if the difference in $\chi^2$ between the first
and second peaks are very close, $\zeta$ increases.  We define three confidence classes based on this statistic.
\begin{equation}
  Q = \left\{
 \begin{array}{l l}
	2 & \mbox{ if log $\zeta > 0.3$} \\
	3 & \mbox{ if $-0.3 < {\rm log} \zeta < 0.3$} \\
	4 & \mbox{ if log $\zeta < -3.0$}\\
\end{array}	\right.\end{equation}
The thresholds for each class are empirically determined by finding values
which maximized sample completeness while minimizing outlier rate. 
With this classification, objects with $Q=4$ have the most certain redshifts.
For most statistical studies, a combination of $Q=3$ and $Q=4$ provides a sample
of galaxies with confident redshifts and low outlier rate. Throughout the rest of this work, 
we utilize a $Q\ge3$ sample when measuring survey completeness. Table \ref{table:zconf} lists the 
scatter and outlier rate for all three quality classes as well as the fraction of the PRIMUS primary galaxy sample
which were classified into each bin.

\subsection{Redshift Success}
\label{sec:redshiftsuccess}
In addition to the science objects observed as part of PRIMUS, we also
observed targets with known redshifts in DEEP2 and VVDS fields for calibration purposes.  After our initial 
observations, zCOSMOS released an early redshift catalog which we further use to create
a sample of PRIMUS objects with known redshifts from higher-resolution spectroscopic 
observations.   In order to quantify the redshift success from our PRIMUS
observations and redshift fitting, we compare our best-fit PRIMUS redshifts to this 
sample of known high-resolution redshifts.

\begin{figure}
\includegraphics[width=3.5in]{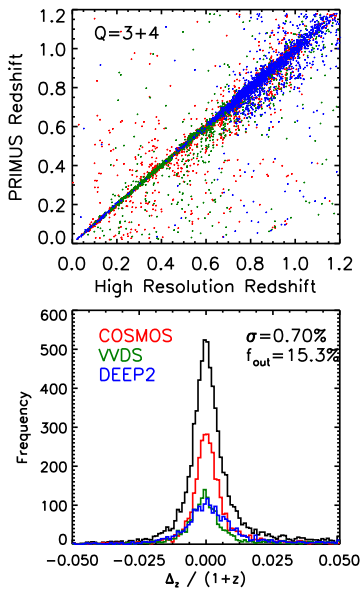}
\caption{\label{fig:zvzq3} Same as Figure \ref{fig:zvzq4} but for galaxies with $Q=3$ and $Q=4$.  The inclusion of $Q=3$ adds some dispersion
compared to the $Q=4$ sample -- the $Q\ge3$ sample shows a $1\sigma$ dispersion of $0.7\%$ and 15.3\% rate of outliers more than $5\sigma$ from
agreement.}
\end{figure}

Figure \ref{fig:zvzq4} shows the best fitting PRIMUS redshift compared to 
higher-resolution spectroscopic redshifts from DEEP2, zCOSMOS, and VVDS for 
$Q=4$ objects and the associated distribution of differences.   Overall we find
excellent agreement between the two redshift measurements with a $\sim0.5\%$ scatter in 
$\Delta z/(1+z)$.  Defining catastrophic outliers as objects with $\Delta z/(1+z)>5\sigma$
we find a total outlier rate of 8\%.    Similarly, 
Figure \ref{fig:zvzq3} shows the same distribution for all $Q=3$ and $Q=4$ galaxies.
We find a 0.7\% scatter rate and 15\% outlier rate.  The $Q\ge3$ sample is the best suited for
statistical studies of galaxy properties as it is more complete.

One of the largest sources of error in PRIMUS redshifts is the presence of degeneracies between
two model spectra at different redshifts.  At PRIMUS resolution, a galaxy at $z>0.7$ with a strong 
4000 \AA\, break can be difficult to differentiate from a bluer galaxy with a Balmer break and a
strong \oii\,. emission line at slightly higher redshift. 
 The addition of high-quality photometry can help break this degeneracy, especially
in the U-band. As an example of the aid that U-band photometry can provide to our fits, in regions with
U-band photometry, we find a 25\% tighter relation between high-resolution redshift and PRIMUS redshift and
the number of outliers located more than $5\sigma$ from the one-to-one relationship dropped by a factor of two.

\section{Survey Completeness}
\label{sec:completeness}
Regardless of the effort invested into any wide-area, deep, galaxy redshift 
survey, 100\% of the objects of interest will not yield redshifts.  
 incompleteness can arise at many phases of the survey process;
galaxies may not be present in photometric catalogs due to catalog depth,
technical limitations or survey time limitations may prevent the 
targeting of every galaxy selected as a possible target from photometric
imaging, and finally observed spectra may fail to provide a reliable redshift
measurement.  Proper use of data obtained from galaxy redshift 
surveys requires the careful correction for these sources
of incompleteness in order to not bias statistical analysis of the sample.
As the photometric catalogs we target from in PRIMUS are often significantly
deeper than the PRIMUS spectroscopic depth, we assume the incompleteness
in the photometric catalogs to be negligible and focus on 
targeting and redshift incompleteness in the PRIMUS observations.
When selecting targets for PRIMUS spectroscopy, we applied two 
{\it a-priori} sparse sampling criteria (see \citet{Coil2010_PaperI} 
for more details).
In order to ensure that the faintest
galaxies, which dominate the photometric catalog, do not dominate the PRIMUS
target selection, we define two bins in flux.  Galaxies more than 0.5 mag
brighter than the flux limit of the PRIMUS selection were sampled at a
rate of 100\% while galaxies below this limit were sampled at a rate of
30\%.  As this sparse sampling is performed during target selection and
the sparse sampling weights were saved, we can correct for the 
magnitude-dependent sparse sampling exactly.   Secondly, PRIMUS
target selection utilized a density-dependent sparse sampling approach to 
ensure that galaxies in (projected) dense regions were not undersampled.  We
can upweight each galaxy by the inverse of the sparse-sampling weight
to reconstruct the full magnitude-limited sample. 

When designing slit masks for PRIMUS observations, not every object selected
as a PRIMUS target can be observed due to slit collisions between PRIMUS 
targets themselves and between targets and higher priority targets such
as alignment stars and flux calibration stars. The effect of these slit 
collisions is minimized by observing two slit masks at each PRIMUS 
field center and by using density-dependent sampling, 
but this does not allow us to observe every possible PRIMUS 
target.   In order to correct for missing targets from slit collisions, 
we calculate the number of  PRIMUS galaxies which were observed and 
divide by the number of PRIMUS galaxies available as a function
of the number of masks each object had the opportunity to be observed on and 
the number of potential conflicting galaxies, $N_c$.  Averaging over all
of the PRIMUS fields except XMM-SXDS, our completeness in areas observed
by two masks is 97\% for $N_C\le2$ and declines with increases $N_C$ to
82\% for $N_c=6$.  In areas covered by four masks (in regions
of the COSMOS field), our average completeness is 98\% for
$1<N_c<4$ and declines slightly to $94\%$ for $N_c=6$.  In XMM-SXDS, 
we observed a factor of two more F-stars per mask than our other fields, which 
leads to lower overall completeness (96\% for $N_c=2$ and 75\% for $N_C=6$).

\begin{figure}
	\includegraphics[width=3.5in]{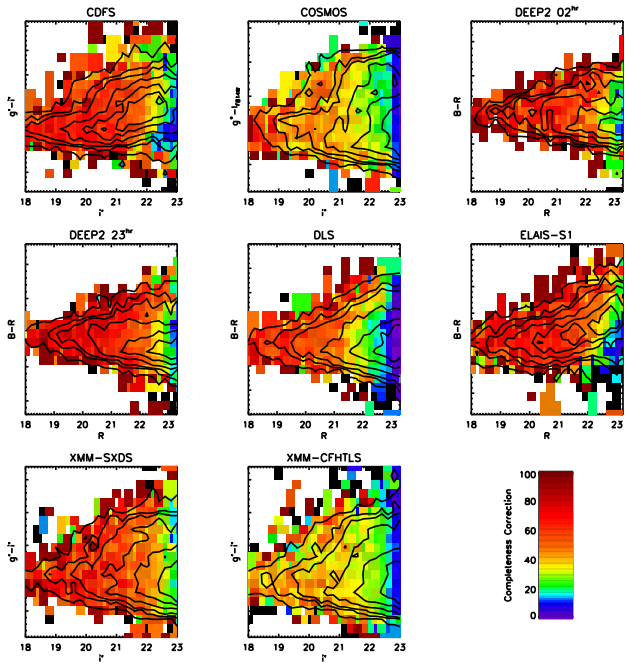}
	\caption{\label{fig:completeness} Redshift success as a function of color and magnitude for each of the PRIMUS fields.  
	Contours enclose 25\%, 50\%, 75\%, 90\%, and 99\% in the targeted PRIMUS sample while the colored background shows the 
	fraction of targeted galaxies with a well-measured ($Q\ge3$) redshift in PRIMUS. 
	Due to the inhomogeneous filter systems utilized
	in the optical imaging across each field, we utilize synthesized $r-i$ color and $i$-band magnitude based on {\tt kcorrect} 
	fits to the observed photometry in this comparison.  Overall, the primary factor in our redshift success rate is the $i$-band 
	flux of the object; little trend is seen with respect to an object's color.  The average
	redshift success is $>75\%$ brighter than $i=21$ and declines to $\sim45\%$ at $i\sim22.5$.}
\end{figure}

Finally, we must correct for galaxies which were observed by PRIMUS but
for which we were unable to measure a reliable (defined as $Q\ge3$, here)
redshift.  In order to quantify this incompleteness and explore its
dependence on galaxy type, we count the number of observed galaxies as
a function of both observed magnitude and color.  As
the photometric systems between each of the imaging datasets used for
target selection used different filter sets, the exact filters used for this comparison
are different between fields.  In general, we choose either the $i$ or $R$ bands 
when quantifying trends with magnitude and choose a color with a large lever arm across the optical
spectrum (typically $g-i$ or $B-R$) when looking at trends with observed-frame color. 
We then derive the final spectroscopic weight in this binned color-magnitude space.  We define
the final weight as the total number of galaxies in a bin with robust redshift measurements
divided by the {\it effective} number of galaxies (that is the total of the PRIMUS statistical 
weights) from the full primary sample in the bin.  
We use this method to ensure that galaxies from populations with few successful redshift measurements
are still given full weight in the final primary sample. Figure \ref{fig:completeness}
shows the completeness for each of the PRIMUS fields.  In each panel
the contours show the concentration of objects in the color-magnitude plane 
and denote 25\%, 50\%, 75\%, 90\%, and 99\% of the population.  The color
denotes the fraction of observed PRIMUS objects which received a 
reliable redshift measurement.  In general, the flux of the
object is the dominant variable when describing the PRIMUS
redshift incompleteness -- little variation is seen in the completeness as a function of color
at fixed flux.  The average redshift success is $>75\%$ for galaxies
brighter than $i=21$ and declines to $\sim 45\%$ at $i=22.5$.

% ####################
\section{Conclusions}
\label{sec:lessons}
PRIMUS has completed observations
of 185,105 galaxies with a low-dispersion prism over 9.1 deg$^2$ 
suitable for statistical investigates of galaxy evolution since
$z\sim1$.  In our primary sample of galaxies with redshift measurements
suitable for statistical work, PRIMUS includes 130,000 galaxies with
typical redshifts errors of $\sigma_z/(1+z)=0.005$ when compared to 
higher-resolution spectroscopic redshift measurements.  PRIMUS is
the largest faint galaxy survey completed to date.  The 
high targeting fraction ($\sim 80\%$) and large survey volume allow
for precise measurements of galaxy properties and large-scale to z$\sim 1$.
In \citet{Coil2010_PaperI}, we summarize the
motivation, observational techniques, target selection, slitmask design, and 
provided details of the PRIMUS observations.  In this paper, we detail the data processing, spectral extraction, 
redshift fitting, and survey completeness of the PRIMUS.

The dramatic increase in multiplexing ability added
when using the  low-dispersion prism technique developed for PRIMUS
compared to more traditional spectroscopic
techniques at higher-resolution allows for a very efficient mapping of large areas of the sky 
using only a fraction of the observing time. The value of high-efficiency spectroscopic follow up will 
increase as the next generation of large area deep sky surveys such as the Dark Energy Survey and LSST
begin providing public data.  The volumes probed by these imaging surveys will be impossible to 
fully spectroscopically probe using traditional techniques.

While the prism technique has many strengths, especially
in terms of survey efficiency, it has several limitations; the effect of these limitations
can be minimized through proper planning and careful survey design. 
As much of the spectral information on small scales (absorption and emission lines features) will be
unresolved with prism spectroscopy, the more information available to break degeneracies in spectral fitting
will result in more precise redshift measurements.  For example, in PRIMUS, a spectral break at $z>0.7$ may 
arise due to a strong 4000 \AA\,  break in a red galaxy or a Balmer break with an unresolved \oii emission line in a star forming 
galaxy.  The addition of external photometry, especially in the $U$-band, helps to breaks this degeneracy.  Having broadband 
photometry which spans the optical spectrum provides significant improvement to redshift fits.

Traditional higher-resolution spectroscopic surveys often utilize sky fibers, extended slits, or nod \& shuffle
procedures to characterize the sky emission in target spectra as the signal from targets of interest are many orders 
of magnitude fainter than the sky and thus any contaminating emission not removed will bias the final extracted spectra. 
This problem is magnified with low dispersion spectroscopy as the entire trace of objects of interest may only be a few hundred
pixels of information. Small scale errors that may affect only a small portion of the spectrum in high-resolution spectroscopic
work (and which can easily be masked out when fitting for redshifts) removes a significant fraction of the spectral 
information at lower-resolution.   Robust sky subtraction is essential, though the most effective technique is strongly dependent 
on the instrument, goals, and depth of each survey. 

A key problem that required significant effort to correct in our PRIMUS observations was contamination of our spectra from large-scale
scattered light in the final images. This scattered light is likely insignificant for the majority of users of the instrument as it exists
on large enough scales to be removed when subtracting sky emission in direct imaging. The scattered light being dependent on the
flux from nearby pixels means that  most traditional high-resolution spectroscopy applications will likely be minimally effected as the counts per pixel
are often significantly smaller than when working with low-dispersion spectra (as the sky emission is resolved rather than being unresolved in the red).  
This underlines the critical importance in understanding the instrument and optical system when performing low-resolution spectroscopic work.   Often
corrections unneeded, or even unnoticed, in more traditional applications will play an important role in the final success of the survey.

Once a characterization of the available photometric data, a coherent plan to remove contaminating emission from object spectra, and a firm understanding of the
instrument and optical systems involved have been established, low-resolution spectroscopic surveys provide very efficient means to observed thousands
of objects simultaneously.  Early success with PRIMUS led to a number of current prism surveys to probe different areas of parameter space including
the high redshift universe, the nature of galaxy cluster members, and the properties of colliding galaxy clusters.

This paper includes data gathered with the 6.5 meter Magellan 
Telescopes located at Las Campanas Observatory, Chile.
We thank the support staff at LCO 
for their help during our observations, and we acknowledge
the use of community access through NOAO observing time.
Some of the data used for this project is from the CFHTLS public data release,
which includes observations obtained with MegaPrime/MegaCam, a joint project 
of CFHT and CEA/DAPNIA, at the Canada-France-Hawaii Telescope (CFHT) which is 
operated by the National Research Council (NRC) of Canada, the Institut 
National des Science de l'Univers of the Centre National de la Recherche 
Scientifique (CNRS) of France, and the University of Hawaii. This work is 
based in part on data products produced at TERAPIX and the Canadian Astronomy 
Data Centre as part of the Canada-France-Hawaii Telescope Legacy Survey, a 
collaborative project of NRC and CNRS.
Funding for PRIMUS has been provided
by NSF grants AST-0607701, 0908246, 0908442,
0908354, and NASA grant 08-ADP08-0019.
RJC was supported by NASA through Hubble Fellowship grant 
HF-01217 awarded by the Space Telescope Science Institute, 
which is operated by the Associated of Universities for
 Research in Astronomy, Inc., for NASA, under contract NAS 5-26555.

\bibliographystyle{apj}
\bibliography{my_bib}

\begin{deluxetable}{cccc}
\tablecaption{PRIMUS Redshift Confidence Classes\label{table:zconf}}
\tablewidth{0pt}
\tablehead{
\colhead{Class} & 
\colhead{$\sigma_{\delta z/(1+z)}$} & 
\colhead{Outliers\tablenotemark{a}} & 
\colhead{Sample Fraction\tablenotemark{b}}
}
\startdata
  4 &    0.005 &       7.85 &       49.2 \\
       3 &     0.022 &       5.32 &       21.6 \\
       2 &     0.050 &       5.06 &       29.2 
\enddata
\tablenotetext{a}{Fraction of objects with known redshifts deviating more than 5$\sigma$ from agreement.}
\tablenotetext{b}{Fraction of PRIMUS primary galaxies which received the specified class designation.}
\end{deluxetable}

\end{document}